\documentclass[runningheads]{llncs}

\usepackage[T1]{fontenc}
\usepackage{cite}
\usepackage{newtxtext}
\usepackage[varvw]{newtxmath}

\usepackage{algorithm}
\usepackage{algpseudocode}
\usepackage{anyfontsize}
\usepackage{bm}
\usepackage{booktabs}
\usepackage{nicefrac}
\usepackage{float}
\usepackage{mathtools}
\usepackage[defaultlines=2, all]{nowidow}
\usepackage{tabularx}
\usepackage{tikz}
\usetikzlibrary{patterns, plotmarks, positioning}
\usetikzlibrary{calc,decorations.pathreplacing}
\usepackage{xcolor}
\usepackage{environ}
\usepackage[dvipsnames]{xcolor}
\usepackage{xurl}
\graphicspath{{graphics/}}

\usepackage{hyperref}
\usepackage{cleveref}

\usepackage{thm-restate}
\newcommand{\I}{\mathcal{I}} 
\DeclareMathOperator{\GF}{GF} 

\newcommand{\R}{\mathbb{R}}

\newcommand{\Z}{\mathbb{Z}}

\renewcommand{\c}{\cdot}

\newcommand{\one}{\mathbf{1}}

\spnewtheorem{observation}{Observation}{\bfseries}{\itshape}
\crefname{observation}{Observation}{Observations}

\newif\ifshowcomments
\showcommentstrue   

\ifshowcomments
  \newcommand{\britta}[1]{\textcolor{rwthmagenta}{\em \small [Britta: #1]}}
  \newcommand{\philipp}[1]{\textcolor{rwthred}{\em \small [Philipp: #1]}}
  \newcommand{\raoul}[1]{\textcolor{rwthpurple}{\em \small [Raoul: #1]}}
  \newcommand{\sam}[1]{\textcolor{ForestGreen}{\em \small [Sam: #1]}}
\else
  \newcommand{\britta}[1]{\ignorespaces}
  \newcommand{\philipp}[1]{\ignorespaces}
  \newcommand{\raoul}[1]{\ignorespaces}
  \newcommand{\sam}[1]{\ignorespaces}
\fi

\newcommand{\U}{\mathcal{U}}

\DeclareMathOperator{\OPT}{OPT}
\DeclareMathOperator{\ALG}{ALG}

\newcommand{\NP}{\textsf{NP}}

\DeclareMathOperator{\XP}{\textsf{XP}}
\DeclareMathOperator{\FPT}{\textsf{FPT}}

\newcommand{\BPPT}{\textsc{BPPT}}
\newcommand{\MWIS}{\textsc{MWIS}}
\newcommand{\IMFT}{\textsc{IMFT}}

\newcommand{\conge}{C}
\newcommand{\ucong}{UC}
\newcommand{\mincap}{\beta}
\newcommand{\UC}{\ucong}
\newcommand{\C}{\mathcal{C}}
\newcommand{\LP}{\textsc{LP}}

\makeatletter
\newif\if@qedplaced
\newcommand{\qedhere}{
  \ifmmode\tag*{\squareforqed}\else\qed\fi
  \global\@qedplacedtrue}
\let\llncs@endproof\endproof
\def\endproof{%
  \if@qedplaced
    \global\@qedplacedfalse
    \let\qed\relax\let\squareforqed\relax
  \fi
  \llncs@endproof}
\makeatother

\makeatletter
\newcommand{\algmargin}{\the\ALG@thistlm}
\makeatother

\algnewcommand{\parState}[1]{\State%
  \parbox[t]{\dimexpr\linewidth-\algmargin}{\strut #1\strut}}

\begin{document}

\title{Path-constrained Packing on Trees}

\newif\ifsubmission
\submissionfalse

\ifsubmission
    \author{Anonymous Authors}
\else
    \author{Samuel Fiorini\inst{1}
    \and
    Raoul Luqué\inst{2}
    \and
    Philipp Pabst\inst{2}
    \and
    Britta Peis\inst{2}
    }
\fi

\ifsubmission
    \authorrunning{Anonymous Authors}
\else
    \authorrunning{S. Fiorini, R. Luqué, P. Pabst, B. Peis}
\fi

\ifsubmission
    \institute{}
\else
    \institute{Université libre de Bruxelles, Bruxelles, Belgium \and RWTH Aachen, Aachen, Germany}
\fi

\maketitle 

\begin{abstract}
An instance of \emph{Budgeted Path-constrained Packing on Trees (\BPPT)} consists of a graph $G$ together with a spanning tree $T$, a weight $w_e \in \Z_{\ge 0}$ for each tree edge $e \in E(T)$ and a budget $b_f \in \Z_{\ge 1}$ for each non-tree edge $f \in E(G) \setminus E(T)$. The goal is to find a maximum weight packing of fundamental cuts respecting the budgets. Equivalently, an instance of \BPPT{} is an integer program of the form $\max \{w^\intercal x : Ax \le b,\ x \in \{0,1\}^n\}$ where $A \in \{0,1\}^{m \times n}$ is a transposed edge-path-tree matrix, $w \in \Z^n_{\ge 0}$ and $b \in \Z^m_{\ge 1}$. We define the \emph{congestion} $\conge$ of a \BPPT{} instance as the maximum number of non-zeros in a column of $A$. The \emph{unit congestion} $\ucong$ is defined similarly, except that rows with $b_i \ge 2$ are ignored. We also consider the minimum budget $\beta := \min \{b_i : i = 1, \ldots, m\}$. In this paper, we begin the study of \BPPT{} through the lens of the three parameters $\conge$, $\ucong$ and $\mincap$, and relative to the standard LP relaxation $\max \{w^\intercal x : Ax \le b,\ x \in [0,1]^n\}$. \BPPT{} generalizes the independent set problem in graphs of maximum degree at most $\conge$ and is a special case of the $b$-matching problem in hypergraphs of rank at most $\conge$.
We obtain approximation algorithms for \BPPT{} with factors $2+\ucong/2$, $1+\conge/2$ and $\mincap/(\mincap-1)$ for $\mincap \geq 2$. These bounds are relative to the standard LP relaxation and, as we show, integrality gap tight with respect to parameters $\ucong$ and $\mincap$, and almost integrality gap tight relative to $\conge$. We complement these results with a $\lceil (\conge+1)/3 \rceil$-approximation for fixed $\conge$ which is not relative to the standard LP relaxation and has $\XP$ running time parameterized by $\conge$.

\end{abstract}

\section{Introduction}

Packing problems form a broad class of combinatorial optimization problems, including many important problems as special cases. In this paper, we think of a packing problem as any restriction of the following problem: given a \emph{constraint matrix} $A \in \Z^{m \times n}_{\ge 0}$, a \emph{right-hand side vector} $b \in \Z^m_{\ge 1}$ and a \emph{weight vector} $w \in \Z_{\ge 0}^n$, solve the integer program
\begin{equation} \label{eq:packing_IP}
    \max \{w^\intercal x : Ax \le b, x \in \{0,1\}^n\}\,.
\end{equation}
For example, the case where $A \in \{0,1\}^{m \times n}$ has two non-zeros per row and $b = \one$ exactly corresponds to the \emph{Maximum Weight Independent Set} problem (\MWIS).
Hence, every hardness result for \MWIS{} transfers to every packing problem that contains all instances $(A,b,w)$ for which $A$ is the edge-vertex incidence matrix of a graph and $b$ is the all-one vector (that is, $b = \one$).
\medskip

\noindent \textbf{Packing integer programs over TU matrices.} Recall that a matrix $A \in \Z^{m \times n}$ is \emph{totally unimodular} (TU) whenever $\det B \in \{-1,0,1\}$ for all square submatrices $B$ of $A$. In case $A$ is TU, the integer program \eqref{eq:packing_IP} can be solved in polynomial time through its \emph{standard LP relaxation}
\begin{equation} \label{eq:packing_LP}
    \max \{w^\intercal x : Ax \le b, x \in [0,1]^n\}\,.
\end{equation}
An important example of TU matrices are \emph{network matrices}. A matrix $A \in \{0, \pm 1\}^{m \times n}$ is said to be a network matrix if there exists a directed graph $G$ with $m + n$ directed edges and a spanning tree $T$ of $G$ with $m$ edges such that the $m$ rows of $A$ correspond to the tree edges $e \in E(T)$ and the $n$ columns of $A$ to the non-tree edges $f \in E(G) \setminus E(T)$, and $A_{ef} = 1$ if $e$ is a backward edge of the fundamental cycle of $f$, $A_{ef} = -1$ if $e$ is a forward edge of the fundamental cycle of $f$, and $A_{ef} = 0$ otherwise.

One reason for the importance of network matrices is Seymour's~\cite{Sey80} decomposition theorem for regular matroids, which translates to a decomposition for TU matrices. Essentially, the result states that every TU matrix can be obtained from network matrices, transposed network matrices, and two special TU matrices, by operations including $1$-, $2$- and $3$-sums, re-signing rows and columns, and pivoting. Among other things, Seymour's decomposition allows us to recognize TU matrices in polynomial time.

Interestingly, much of the theory of regular matroids is done over $\GF(2)$, where $-1 = 1$. Furthermore, most TU recognition algorithms forget the signs in the input matrix as their very first step, see e.g., \cite{Tru90, WalTru92}. This indicates the relevance of the concept for a binary matrix to ``be TU with a correct signing of the entries''. More formally, a matrix $A \in \{0,1\}^{m \times n}$ is said to be \emph{TU-signable} if it is the support matrix of a TU matrix $A^\pm \in \{-1,0,1\}^{m \times n}$. A binary matrix $A$ is TU-signable if and only if the vector matroid $M[I\ A]$ is regular, when computed over $\GF(2)$. It is known that the TU signing of a TU-signable matrix is unique, up to re-signing whole rows and columns, and can be obtained via Camion's signing algorithm \cite{Camion1965}. In light of Seymour's decomposition, the most important classes of TU-signable matrices are the supports of network matrices and their transposes. We call these matrices (transposed) \emph{edge-path-tree} (EPT) matrices.

Formally, $A \in \{0,1\}^{m \times n}$ is an EPT matrix if there exists a tree $T$ with $m$ edges $e_1$,~\ldots,~$e_m$, and $n$ subpaths $P_1$, \ldots, $P_n$, such that $A_{ij} = 1$ if $e_i$ is an edge of $P_j$, and $A_{ij} = 0$ otherwise. In particular, every $m \times n$ binary matrix $A$ with two non-zeros per column is EPT. This can be seen by taking $T$ to be a star with $m$ edges corresponding to the rows of $A$ and considering a leaf-to-leaf path $P_j$ for each column of $A$.
\medskip

\noindent \textbf{Packing integer programs over TU-signable matrices.} For understanding packing problems over TU-signable matrices, it is important to understand packing problems over the building blocks of TU-signable matrices, namely, EPT matrices and transposed EPT matrices. Packing integer programs \eqref{eq:packing_IP} over EPT matrices were studied by Garg, Vazirani and Yannakakis~\cite{GVY97}. The corresponding problem is known as the \emph{maximum Integral Multicommodity Flow problem in Trees (IMFT)}. Garg \emph{et al.} considered the unweighted case (that is, $w = \one$), proved that IMFT can be solved in polynomial time whenever $b = \one$, gave a $2$-approximation for general right-hand side vectors $b$, and proved that the problem is APX-hard for general $b$.
For the weighted case, the best known result is a $4$-approximation by Chekuri, Mydlarz and Shepherd~\cite{Chekuri07}.

In this work, we focus on packing integer programs \eqref{eq:packing_IP} over \emph{transposed} EPT matrices. In other words, we consider the following problem. \\
\vspace{-0.15cm}

\ifsubmission
\else
\newpage
\fi

\noindent\textsc{Budgeted Path-constrained Packing on Trees (\BPPT)}
\vspace{-0.15cm}
\begin{description}
  \item[Input:] Tree $T$, \emph{commodities} $(P_i, b_i)$ for $i \in [m]$, each consisting of a path $P_i \subseteq T$ and a \emph{budget} $b_i \in \Z_{\geq 1}$, and edge weights $w_e \in \Z_{\geq 0}$ for all $e \in E(T)$.
  \item[Goal:] Find edge set $F \subseteq E(T)$ with $|F \cap E(P_i)| \leq b_i$ for all $i \in [m]$, maximizing $w(F) \coloneqq \sum_{e \in F} w_e$.
\end{description}

If $T$ is a path, then the constraint matrix $A$ has the consecutive-ones property. More generally, we call an instance of \BPPT{} \emph{vertical} if $T$ can be rooted in a way such that each path $P_i$ is \emph{vertical}, that is, runs from a vertex to one of its ancestors. In this case, $A^\intercal$ is a network matrix and \BPPT{} can be solved in polynomial time. Fix an arbitrary root for $T$. We say that a path $P_i$ is \emph{bent} whenever it is not vertical. We define the \emph{apex} of a path $P_i$ as the least common ancestor of its ends. Hence, a path is bent iff its apex lies in its interior. It is the bent paths that make \BPPT{} hard.

We define the \emph{congestion} $\conge$ of a \BPPT{} instance as the maximum over $e \in E(T)$ of the number of commodities $(P_i,b_i)$ such that $e \in E(P_i)$. As we will see later, commodities for which $b_i = 1$ are much harder to deal with algorithmically than commodities for which $b_i \geq 2$. This leads to the notion of \emph{unit congestion} $\ucong$, which we define as the maximum over $e \in E(T)$ of the number of commodities $(P_i,b_i)$ such that $e \in E(P_i)$ and $b_i = 1$. Observe that $\conge$ is the maximum number of non-zeros in a column of the constraint matrix $A$, and similarly for $\ucong$, provided that rows for which $b_i \ge  2$ are ignored.
Clearly, we have $0 \leq \ucong \leq \conge$. The gap between these two parameters is, in general, unbounded. Finally, we define $\mincap$ as the smallest budget $b_i$, where $i \in [m]$. Notice that $\ucong = 0$ iff $\mincap \ge 2$.

The restriction of \BPPT{} to stars $T$ (and $b = \one$) precisely coincides with MWIS on general graphs $G$. For such \BPPT{} instances, $\conge = \ucong = \Delta$, where $\Delta = \Delta(G)$ denotes the maximum degree of the underlying graph $G$.
When considering \eqref{eq:packing_IP} for \emph{any} $A \in \{0,1\}^{m \times n}$, we obtain the \emph{maximum weight $b$-matching problem} in a hypergraph $H=(V,E)$ with $|V| = m$ and $|E| = n$. Let $k$ denote the \emph{rank} of this hypergraph, that is, the maximum number of vertices in a hyperedge. Notice that $k$ is also the maximum number of non-zeros in a column of $A$. Hence, if $A$ is a transposed EPT matrix, then $k = \conge$ where $\conge$ denotes the congestion of the corresponding \BPPT{} instance. 

To sum up the last paragraph, we see that \BPPT{} ``sits between'' \MWIS{} (in graphs of maximum degree $\conge$) and maximum weight $b$-matching (in hypergraphs of rank $\conge$).

\subsection{Related Work}

To the best of our knowledge, there is no prior work on \BPPT. However, the problem is closely related to several well-studied problems other than \MWIS{} and weighted hypergraph $b$-matching.\medskip

\noindent \textbf{Maximum weight independent set.} It is known that approximating \MWIS{} within an $n^{1-\varepsilon}$-factor is \NP-hard for every positive $\varepsilon$, where $n$ denotes the number of vertices of the graph \cite{Zuckerman06}. Moreover, approximating \MWIS{} within an $\mathcal{O}(\Delta /\log^2 \Delta)$-factor is also \NP-hard, assuming the unique games conjecture, see~\cite{AustrinKhotSafra11inapproximabilty}. These strong inapproximability results are inherited by \BPPT{}. 
There is a long line of work on approximating \MWIS{} in bounded-degree graphs. The first major result in this line is a $\Delta/2$-approximation by Hochbaum~\cite{Hochbaum83boundedDegreeIS} combining polyhedral properties of the standard LP relaxation, namely, half-integrality and persistency~\cite{nemhauserTrotter75halfIntegrality}, together with coloring arguments, namely, Brooks' theorem. Halldórsson and Lau \cite{HalldorssonLau97} improved this to a $(\Delta + 2)/3$-approximation using a partition of the input graph into sufficiently few subgraphs of maximum degree $2$. 

In this work we aim to extend these results and techniques to approximation algorithms for \BPPT{}, parameterized by one of $\conge$, $\ucong$ and $\mincap$. We point out that \BPPT{} generalizes \MWIS{} in two distinct directions, both of which pose nontrivial challenges. First, \MWIS{} captures \BPPT{} instances on stars, whereas we consider general trees throughout this work. Second, in \BPPT{} we have to deal with arbitrary budgets $b_i$ and not only with the unit-budget case.

\MWIS{} does admit approximation algorithms with approximation factors that are sublinear in $\Delta$. For instance, an $\mathcal{\tilde{O}}(\Delta / \log \Delta)$-approximation was achieved both by Halldórsson \cite{Halldorsson00} and Halperin \cite{Halperin02}, and an $\mathcal{\tilde{O}}(\Delta / \log^2 \Delta)$-approximation was achieved by Bansal, Gupta and Guruganesh~\cite{BGG18}, which matches the hardness of the problem (ignoring factors that are polynomial in $\log \log \Delta$). We leave the question of adapting these algorithms to \BPPT{} open for future work.\medskip

\noindent \textbf{Weighted hypergraph $b$-matching.} The special case of maximum weight $b$-matching in rank-$k$ hypergraphs with $b = \one$ is also known as the \emph{weighted $k$-set packing problem}, and
is one of the most well-studied packing problems in combinatorial optimization. 
A long line of work on local search algorithms \cite{BafnaNR96, ArkinH98, Berman00, ChandraH01, Neuwohner21, Neuwohner23SetPacking1, Neuwohner23SetPacking2, ThieryWard23} has led to approximation guarantees slightly below $(k+1)/2$. However, such guarantees come at the expense of running time, since all of them are obtained using local exchanges of $k$ or more sets, which takes time $n^{\Omega(k)}$. Such algorithms have polynomial complexity only if $k$ is a constant. Local search with constant-size local exchanges gives a truly polynomial-time approximation algorithm with guarantee close to $0.65k$~\cite{BermanKrysta03constantSizeLocalSearch}. Weighted $k$-set packing is known to be \NP-hard to approximate within $\mathcal{O}(k/\log k)$; this holds even for the unweighted version~\cite{Hazan2006}.

For general right-hand side vectors $b$, maximum weight $b$-matching in rank-$k$ hypergraphs admits a $(k-1+1/k)$-approximation relative to the standard LP relaxation~\cite{ChanLau12LPRounding,weighted_b_matching_rank_k_hypergraphs_apx}. The $b$-matchings of any rank-$k$ hypergraph form a \emph{$k$-extendible system} in the sense of Mestre~\cite{mestre06extendible}. We show that they also form a \emph{$k$-exchange system} in the sense of Feldman, Naor, Schwartz and Ward~\cite{feldman11kexchange}, see 
Appendix \ref{app:kexchange}. From this, we conclude that the greedy algorithm is a $k$-approximation algorithm, and that the non-oblivious local search of Feldman \emph{et al.}~\cite{feldman11kexchange} is a $((k+1)/2+\varepsilon)$-approximation algorithm, at least whenever $k$ is constant.

Since \BPPT{} is a special case of weighted $b$-matching in hypergraphs of rank $k = \conge$, the algorithms mentioned above apply directly to \BPPT. Compared to general instances of hypergraph $b$-matching, for \BPPT{} we can make use of the underlying tree structure to design improved approximation algorithms. Indeed, we will see that this has strong algorithmic consequences.\medskip

\noindent\textbf{Covering integer programs over TU-signable matrices.} Flipping the inequalities in \eqref{eq:packing_IP} and setting $b := \one$, we obtain \emph{covering} integer programs $\min \{c^\intercal x : Ax \ge \one,\ x \in \{0,1\}^n\}$. When $A$ is an EPT matrix, this is the \emph{Weighted Tree Augmentation Problem (WTAP)}, see e.g.\ \cite{CheriyanKKK08, FioriniGKS18}. When $A$ is a transposed EPT matrix, the problem is known as the \emph{weighted multicut problem in trees}, see~\cite{GVY97}. Both problems admit a $2$-approximation relative to the standard LP relaxation, and it is natural to ask whether this holds for all TU-signable matrices, see~\cite[Conjecture 5.5.3]{ZlatinPhD24}. To our knowledge, this question is still open. Traub and Zenklusen obtained better approximation factors for WTAP, culminating in a recent $(3/2+\varepsilon)$-approximation, see~\cite{TZ25,TZSODA22}. \Cref{tab:TU-signable_IPs} summarizes the state of the art on integer programs over TU-signable matrices. \medskip
\begin{table}[h]
    \centering
    \begin{tabular}{|r|c|c|}
\hline
&Packing ($Ax \le b$) &Covering ($Ax \ge \one$)\\
\hline
\hline
$A$ EPT & 
\begin{minipage}{5.8cm}
\vspace{1ex}
\emph{Name:} \IMFT\\
\emph{Star case:} Weighted $b$-Matching\\
\emph{Best apx:} $2$ unweighted \cite{GVY97}, $4$ weighted \cite{Chekuri07}
\vspace{1ex}
\end{minipage}
& 
\begin{minipage}{5cm}
\vspace{1ex}
\emph{Name:} WTAP\\
\emph{Star case:} Weighted Edge Cover\\
\emph{Best apx:} $3/2+\varepsilon$ \cite{TZSODA22}
\vspace{1ex}
\end{minipage}\\
\hline
$A^\intercal$ EPT & 
\begin{minipage}{5.8cm}
\vspace{1ex}
\emph{Name:} $\BPPT$ \ \textbf{(this work)}\\
\emph{Star case:} \MWIS{}\\
\emph{Best apx:} $2+\ucong/2$, $\mincap/(\mincap-1)$, $1 + \conge/2$
\vspace{1ex}
\end{minipage}
& 
\begin{minipage}{5cm}
\vspace{1ex}
\emph{Name:} Weighted Multicut in Trees\\
\emph{Star case:} Weighted Vertex Cover\\
\emph{Best apx:} $2$ \cite{GVY97}
\vspace{1ex}
\end{minipage}
\\
\hline
\end{tabular}

\vspace*{1ex}

\caption{Summary of known results on (packing or covering) integer programs over TU-signable matrices. The contributions of this paper can be found in the bottom-left cell.}
\vspace*{-0.55cm}
\label{tab:TU-signable_IPs}
\end{table}

\noindent \textbf{Integrality gaps.} In this work, we consider integrality gaps of relaxations relative to a single parameter. For instance, consider the standard LP relaxation \eqref{eq:packing_LP}. Let $\OPT$ and $\OPT_{\text{LP}}$ respectively denote the optimum values of \eqref{eq:packing_IP} and \eqref{eq:packing_LP}. The \emph{integrality gap} of \eqref{eq:packing_LP} relative to some parameter $\pi$ is computed as the supremum of $\OPT_{\text{LP}} / \OPT$ over all instances sharing the same value of $\pi$.
It turns out that all the approximation algorithms mentioned above for \MWIS{} are relative to relaxations that are stronger than the standard LP relaxation. Still, Hochbaum's approach implies that the integrality gap of \eqref{eq:packing_LP} relative to the maximum degree $\Delta$ is $(\Delta + 1)/2$. 
The algorithms with sublinear approximation factors mentioned above all use positive semidefinite relaxations.

Finally, we discuss integrality gaps for maximum weight $b$-matching in rank-$k$ hypergraphs. It is known that the integrality gap of the standard LP relaxation \eqref{eq:packing_LP} is precisely $k - 1 + 1/k$~\cite{weighted_b_matching_rank_k_hypergraphs_integrality_ratio}. 
At least for $b = \one$, there exist convex relaxations with integrality gap at most $(k+1)/2$~\cite{ChanLau12LPRounding}. \medskip

\noindent\textbf{Zoning problems in transport networks.} Our problem can be interpreted as a zoning problem as follows: The commodities $(P_i,b_i)$ for $i \in [m]$ are streams of customers that want to travel along fixed paths $P_i$ in graph $G$. Each of these streams has a known willingness-to-pay that determines the budget $b_i$. Furthermore, the weights $w_e$ correspond to the number of customers traversing edge $e$. The goal is to cut the graph into fare zones in order to maximize revenue. Similar models have been studied extensively from a more practical perspective \cite{BabelKellerer03fareZones, HamacherSchoebel04fareZones, OttoBoysen17fareZones}. From a theoretical perspective, a similar model has been studied by Hoefer \emph{et al.} \cite{hoefer26FZA}, who give a sublogarithmic approximation on trees for their problem variant. In contrast to \BPPT, their model does not treat the budgets $b_i$ as hard constraints, which significantly changes the nature of the problem. Instead, they allow solutions that violate the willingness-to-pay of some commodities, in which case the corresponding customers do not generate any revenue. \medskip

\noindent\textbf{Fractional local ratio.} The main tool behind our result in \Cref{sec:2_plus_uc_half} is the \emph{fractional local ratio} technique. Its core idea is to use an optimal LP-solution to iteratively select elements whose local neighborhood carries little LP-mass. This technique was introduced by Bar-Yehuda \emph{et al.} in the context of an interval scheduling problem \cite{BarYehudaHNSS06fractional} and later used by Chan and Lau \cite{ChanLau12LPRounding} for hypergraph matching and by Parekh and Pritchard \cite{weighted_b_matching_rank_k_hypergraphs_apx} for hypergraph $b$-matching. It builds on the older \emph{local ratio} technique, which goes back to Bar-Yehuda and Even \cite{BarYehudaE85}. The technique was later used, for instance, as a unified framework for scheduling problems by Bar-Noy \emph{et al.} \cite{BarNoyBFNS01} and by Ye and Borodin in the context of \MWIS~\cite{YeB12}. For a comprehensive survey, see~\cite{BarYehudaBFR04}.\medskip

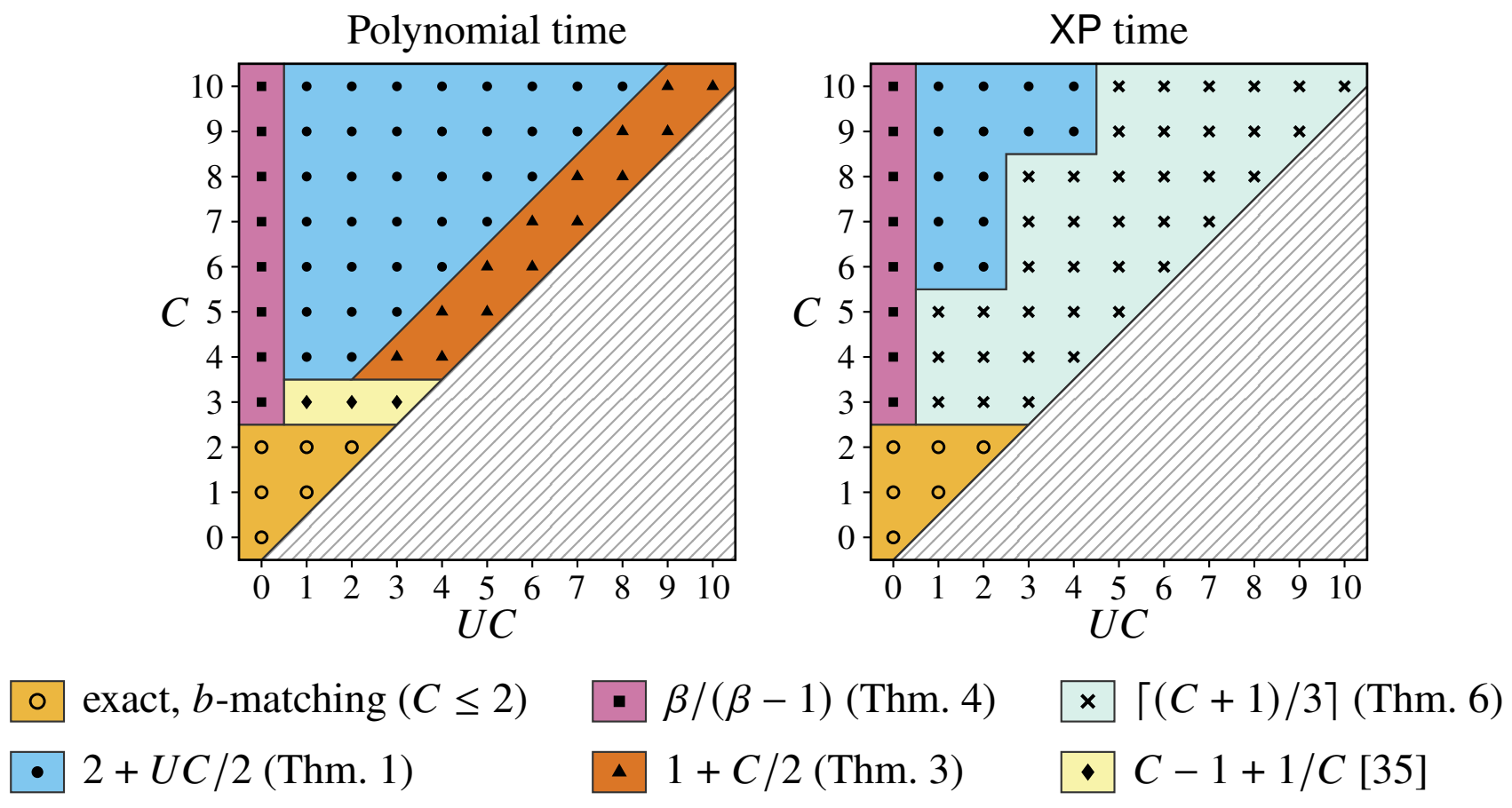
\begin{figure}[t]
\centering

\definecolor{bpptExact}{HTML}{E69F00}
\definecolor{bpptBeta}{HTML}{CC79A7}
\definecolor{bpptUC}{HTML}{56B4E9}
\definecolor{bpptC}{HTML}{D55E00}
\definecolor{bpptHL}{HTML}{009E73}
\definecolor{bpptPP}{HTML}{F0E442}
\tikzset{
  bppt exact/.style   = {fill=bpptExact!75},
  bppt beta/.style    = {fill=bpptBeta},
  bppt uc/.style      = {fill=bpptUC!75},
  bppt c/.style       = {fill=bpptC!85},
  bppt hl/.style      = {fill=bpptHL!15},
  bppt pp/.style      = {fill=bpptPP!45},
  bppt mk exact/.style = {mark=o, mark size=1.5pt,
                          mark options={line width=0.55pt}},
  bppt mk beta/.style  = {mark=square*, mark size=1.25pt},
  bppt mk uc/.style    = {mark=*, mark size=1.3pt},
  bppt mk c/.style     = {mark=triangle*, mark size=1.8pt},
  bppt mk hl/.style    = {mark=x, mark size=1.9pt,
                          mark options={line width=0.75pt}},
  bppt mk pp/.style    = {mark=diamond*, mark size=1.9pt},
  bppt none/.style    = {pattern=north east lines, pattern color=black!35},
  bppt border/.style  = {draw=black!80, line width=0.4pt},
  every mark/.append style={scale=0.8},
}

\newcommand{\bpptRegion}[2]{%
  \path[bppt #1, bppt border] #2;
  \begin{scope}
    \clip #2;
    \foreach \ux in {0,...,10}{\foreach \cy in {\ux,...,10}{%
      \draw plot[only marks, bppt mk #1] coordinates {(\ux,\cy)};}}
  \end{scope}}

\newcommand{\bpptFrame}[1]{%
  \path[bppt none] (0,-.5) -- (10.5,-.5) -- (10.5,10) -- cycle;
  \draw[bppt border] (0,-.5) -- (10.5,10);
  \draw[line width=0.5pt] (-.5,-.5) rectangle (10.5,10.5);
  \foreach \tk in {0,...,10}{%
    \draw[line width=0.4pt] (\tk,-.5) -- ++(0,-2pt)
      node[below, inner sep=1.5pt, font=\scriptsize] {$\tk$};
    \draw[line width=0.4pt] (-.5,\tk) -- ++(-2pt,0)
      node[left, inner sep=1.5pt, font=\scriptsize] {$\tk$};}
  \node[below, inner sep=1pt] at (5,-1.45) {$\ucong$};
  \node[left, inner sep=1pt]  at (-1.45,5) {$\conge$};
  \node[anchor=base, font=\small] at (5,10.95) {#1};
}

\begin{tikzpicture}[xscale = .8, yscale = .8, x=0.42cm, y=0.42cm]
  \begin{scope}
    \bpptRegion{exact}{(-.5,-.5) -- (0,-.5) -- (3,2.5) -- (-.5,2.5) -- cycle}
    \bpptRegion{beta}{(-.5,2.5) rectangle (.5,10.5)}
    \bpptRegion{pp}{(.5,2.5) -- (3,2.5) -- (4,3.5) -- (.5,3.5) -- cycle}
    \bpptRegion{c}{(2,3.5) -- (4,3.5) -- (10.5,10) -- (10.5,10.5) -- (9,10.5)
      -- cycle}
    \bpptRegion{uc}{(.5,3.5) -- (2,3.5) -- (9,10.5) -- (.5,10.5) -- cycle}
    \bpptFrame{Polynomial time}
  \end{scope}
  \begin{scope}[shift={(14,0)}]
    \bpptRegion{exact}{(-.5,-.5) -- (0,-.5) -- (3,2.5) -- (-.5,2.5) -- cycle}
    \bpptRegion{beta}{(-.5,2.5) rectangle (.5,10.5)}
    \bpptRegion{hl}{(.5,2.5) -- (3,2.5) -- (10.5,10) -- (10.5,10.5)
      -- (4.5,10.5) -- (4.5,8.5) -- (2.5,8.5) -- (2.5,5.5) -- (.5,5.5) -- cycle}
    \bpptRegion{uc}{(.5,5.5) -- (2.5,5.5) -- (2.5,8.5) -- (4.5,8.5)
      -- (4.5,10.5) -- (.5,10.5) -- cycle}
    \bpptFrame{$\XP$ time}
  \end{scope}
\end{tikzpicture}

\medskip

\newcommand{\bpptKey}[1]{%
  \tikz[baseline={([yshift=-0.25em]current bounding box.center)}]{%
    \path[bppt #1, bppt border] (0,0) rectangle (1.25em,0.95em);
    \draw plot[only marks, bppt mk #1] coordinates {(0.625em,0.475em)};}\,}
{\footnotesize
\begin{tabular}{@{}l@{\hspace{1.5em}}l@{\hspace{1.5em}}l@{}}
  \bpptKey{exact} exact, $b$-matching ($\conge \le 2$)
  & \bpptKey{beta} $\mincap/(\mincap-1)$
      (Thm.~\ref{thm:ell_ell_minus_one_approx})
  & \bpptKey{hl} $\lceil(\conge+1)/3\rceil$
      (Thm.~\ref{thm:congestion_approx_halldorsson_lau})
  \\[4pt]
  \bpptKey{uc} $2 + \ucong/2$
      (Thm.~\ref{thm:two_plus_uc_half_approx})
  & \bpptKey{c} $1+\conge/2$
      (Thm.~\ref{thm:flrCongestionImproved})
  & \bpptKey{pp} $\conge-1+1/\conge$
      \cite{weighted_b_matching_rank_k_hypergraphs_apx}
\end{tabular}
}

\caption{Best known approximation guarantees among our algorithms and the work of Parekh and Pritchard \cite{weighted_b_matching_rank_k_hypergraphs_apx} for different pairs $(\conge, \ucong)$ with $\conge \geq \ucong$. On the left-hand side we only consider polynomial-time algorithms, whereas on the right-hand side we also include the coloring algorithm from \Cref{thm:congestion_approx_halldorsson_lau} whose running time is in $\XP$ parameterized by $\conge$.}
\vspace*{-0.2cm}
\label{fig:best_algorithm_regions}
\end{figure}

\subsection{Our Results} \label{sec:results}

Our two main results are approximation algorithms relative to the standard LP relaxation \eqref{eq:packing_LP}, with approximation guarantees $2+\ucong/2$ (\Cref{thm:two_plus_uc_half_approx}) and  $\mincap/(\mincap-1)$ (\Cref{thm:ell_ell_minus_one_approx}). We show that each one of these guarantees exactly matches the integrality gap relative to the corresponding parameter, see~\Cref{thm:two_plus_uc_half_integrality_gap,thm:ell_ell_minus_one_integrality_gap}.

Both algorithms build an auxiliary instance $\mathcal{I}'$ that possesses extra structural properties but at the same time is still sufficiently close to the initial instance $\mathcal{I}$. The general idea is to pick an arbitrary root and split certain commodities into two or more commodities, making sure that for split commodities all resulting pieces are vertical and using an optimal fractional solution $x^*$ as a guide. In the $(2+\ucong/2)$-approximation, the auxiliary instance crucially only contains unit-budget commodities. Each commodity $(P_i,b_i)$ with $b_i \ge 2$ is split in such a way that every piece $(P,1)$ has $x^*(E(P)) := \sum_{e \in E(P)} x^*_e$ at most $2$, and at least $1$ for all pieces except the last on each side. In the $\mincap / (\mincap - 1)$-approximation, we seek to obtain a vertical instance. Each bent commodity $(P_i,b_i)$ is split into two halves such that each half $(P,b)$ has budget $b \in \mathbb{Z}_{\ge 1}$ roughly proportional to $x^*(E(P))$, where the scaling factor is no worse than $(\beta-1)/\beta$.

\Cref{thm:two_plus_uc_half_approx} improves on the $(\conge+1)/2$-guarantee of non-oblivious local search \cite{feldman11kexchange} whenever $\conge > \ucong + 3$ and avoids the exponential dependence on $\conge$ in the running time of that algorithm. In terms of $\conge$ alone, a variant of the algorithm used to prove \Cref{thm:two_plus_uc_half_approx} with a slightly more involved construction of the auxiliary instance yields an LP-relative approximation guarantee of $1 +\conge/2$ in polynomial time, see Appendix~\ref{app:flr_parameter_congestion_improved}. This improves over \Cref{thm:two_plus_uc_half_approx} in case $\conge \leq \ucong + 1$, and puts the integrality gap of the standard LP relative to $\conge$ in the interval $[1/2 + \conge/2, 1 + \conge/2]$.

In case the minimum budget $\mincap$ is at least $2$, we have $\ucong = 0$ and \Cref{thm:two_plus_uc_half_approx} gives a $2$-approximation. \Cref{thm:ell_ell_minus_one_approx} states that the approximation guarantee can be improved further when $\mincap \geq 3$. For $\mincap = 2$, this result recovers the guarantee of \Cref{thm:two_plus_uc_half_approx} with a much simpler algorithm, and improves on it if $\mincap \ge 3$. 

Finally, in Appendix~\ref{sec:coloring_approx}, we consider $\BPPT$ when parameterized by $\conge$ and extend the $\lceil (\Delta+1)/3 \rceil$-approximation for \MWIS{} by Halldórsson and Lau \cite{HalldorssonLau97} to our problem. However, when dealing with arbitrary trees $T$ instead of only stars, this needs additional runtime, and our algorithm only runs in $\FPT$-time for $b = \one$ and in $\XP$-time for general $b$.
We refer to \Cref{fig:best_algorithm_regions} for an overview that shows which of our algorithms is best for different combinations of parameters $(\conge, \ucong)$.

\section{A $\left(2 + \frac{\UC}{2}\right)$-approximation Algorithm}
\label{sec:2_plus_uc_half}

In this section we show that there exists an efficient $(2 + \UC/2)$-approximation algorithm for $\BPPT$ which is integrality gap tight. That is, we prove the following two theorems.

\begin{restatable}{theorem}{TwoPlusUCHalfApx}
\label{thm:two_plus_uc_half_approx}
    There exists a polynomial-time algorithm that, given a \BPPT{} instance, computes a feasible solution of value at least $\OPT_{LP}/(2+\ucong/2)$.
\end{restatable}
\vspace{-2ex}
\begin{restatable}{theorem}{TwoPlusUCHalfIntegralityGap}
\label{thm:two_plus_uc_half_integrality_gap}
    For each fixed $\ucong \ge 0$, the integrality gap of the standard LP relaxation of \BPPT{} is exactly $2+\ucong /2$.
\end{restatable}

Our algorithm first constructs an auxiliary instance $\I'$ with unit budgets by splitting each bent commodity with $b_i \geq 2$ at its apex, and cutting up each vertical commodity in the resulting instance for which  $b_i \geq 2$ into smaller pieces guided by an optimal LP solution. The main tool for our analysis in \Cref{sec:2_plus_uc_half_algo_analysis} is \Cref{lem:cheap_neighborhood_edge}, which shows that in the resulting auxiliary instance we can iteratively find edges $e$ with little remaining LP weight in their neighborhood. This allows us to apply a common fractional local ratio argument (see e.g. \cite{BarYehudaHNSS06fractional}) which yields \Cref{thm:two_plus_uc_half_approx}. We describe the construction used to prove \Cref{thm:two_plus_uc_half_integrality_gap} in \Cref{sec:two_uc_half_integrality_gap}, and then show that the constructed instance actually attains the claimed integrality gap in Appendix \ref{appendix:two_plus_uc_half_integrality_gap}. Lastly, we show that a variant of the presented algorithm with a more involved construction of the auxiliary instance gives a $(1 + \conge/2)$-approximation when parameterized by $\conge$ (cf. Appendix~\ref{app:flr_parameter_congestion_improved}).

\subsection{Description of the Algorithm}\label{sec:2_plus_uc_half_algo_descr}

Given an instance $\I$ of \BPPT, in its first phase, the algorithm constructs an auxiliary instance $\I'$ consisting solely of unit-budget commodities as follows:
It starts by computing an optimal solution $x^*$ of the LP relaxation. The algorithm then selects an arbitrary vertex $r$ as the root of the underlying tree $T=(V,E)$. This partitions the commodities into vertical and bent commodities. Every bent commodity with $b_i \geq 2$ is split at its apex into a left path and a right path, with the same budget as the original commodity. Afterwards, each vertical commodity in the resulting instance with $b_i \geq 2$ is split further via the following iterative procedure guided by $x^*$, starting at its apex: walk down the path (away from the root), while summing up the $x^*$-values of the edges on the path. Whenever the value of the sum would exceed $2$ when adding the $x^*$-value of the next edge, the current path is split into two at the common vertex of the current and next edge. This procedure is repeated with the remaining path starting at the splitting vertex, and all resulting paths are assigned a budget of $1$. In the resulting instance $\I'$, all commodities have a budget of $1$, and the only bent commodities are those corresponding to bent, unit-budget commodities of the original instance. We refer to the vertical commodities of $\I'$ obtained by splitting up the non-unit-budget commodities of $\I$ as \emph{pieces}.

In its second phase, the algorithm constructs an integral solution $F_{\ALG}$ for $\I$ via the following iterative procedure: Initialize $t\coloneqq 0$, $w_0 := w$,  $\I'_0:= \I'$,   $y:=\frac{x^*}{2}$, and $F:=\emptyset$. In iteration $t$, denote by $N_{\I'_t}[e] = \{ e \} \cup \{ f \in E(\I'_t) \mid e \text{ and } f \text{ share a commodity in $\I'_t$}\}$
the neighborhood of edge $e$ in instance $\I'_t$, where $E(\I'_t)$ denotes the set of (tree) edges of instance $\I'_t$. While the current instance $\I'_t$ contains at least one edge, the algorithm selects an edge $e_t$ minimizing $y(N_{\I'_t}[e])$ over $E(\I'_t)$, adds $e_t$ to $F$, and updates the weights to $w_{t+1}(e):=w_t(e)-w_t(e_t)$ for all edges $e$ in the neighborhood $N_{\I'_t}[e_t]$, and to $w_{t+1}(e):=w_t(e)$ if $e$ does not belong to $N_{\I'_t}[e_t]$. 
Next, the algorithm constructs an instance $\I'_{t+1}$ by contracting all edges $f$ in $\I'_{t}$ for which $w_{t+1}(f) \leq 0$, and increments $t$ to $t+1$. 
Once $\I'_t$ contains no more edges, the algorithm constructs the solution $F_{\ALG}$ using a \emph{reverse greedy} approach. Initialize $F_{\ALG} \coloneqq \emptyset$ and iterate over the selected edges $F = \{e_0, \ldots, e_s\}$ in reverse order and add them to $F_{\ALG}$ if the edge set remains feasible for $\I'$.
Pseudocode of the algorithm is shown in Appendix \ref{appendix:2_UC_half_pseudocode}.

\subsection{Analysis}\label{sec:2_plus_uc_half_algo_analysis}

The key to our analysis is \Cref{lem:cheap_neighborhood_edge}, which allows us to bound the $y$-mass of the neighborhood $N_{\I'_t}[e_t]$ in each iteration $t$ by $1+\frac{UC}{4}$. Using this fact, we can apply the fractional local ratio framework \cite{BarYehudaHNSS06fractional} to prove the desired approximation guarantee for the constructed solution $F_{\ALG}$ to the auxiliary instance $\I'$. By \Cref{lem:properties_i_prime}, this solution is also feasible for the original instance $\I$. We prove \Cref{lem:cheap_neighborhood_edge} in detail in this section, and defer the proof of the approximation guarantee, that is of \Cref{thm:two_plus_uc_half_approx}, to Appendix~\ref{appendix:proof_two_uc_half_apx}.

\begin{observation}\label{lem:properties_i_prime}
Consider the auxiliary instance $\I'$ and $y := x^*/2$ defined above. Then,
\begin{enumerate}
    \item[(a)] Every feasible solution for $\I'$ is feasible for the original instance $\I$.
    \item[(b)] For every commodity of $\mathcal{I}'$ with path $P$, we have $y(P) \le 1$ if $P$ is vertical, and $y(P) \le \tfrac{1}{2}$ if $P$ is bent.
\end{enumerate}
\end{observation}
\begin{proof}
    We show that the algorithm splits each commodity $(P_i,b_i)$ of the original instance into at most $b_i$ pieces when constructing $\I'$, which immediately implies statement (a).  We consider both sides of $P_i$ separately. If a side of $P_i$ is split into $k \geq 2$ pieces, we know that each of them except for possibly the last one has LP mass at least 1. Additionally, the last two pieces of the side have LP mass strictly larger than $2$ as otherwise the algorithm would not have made a cut between the last two pieces. Together, this shows that if a side is cut into $k \geq 2$ pieces, the LP weight of that side must be at least $k$. When summing over both sides of a commodity, we thus obtain at most $b_i$ pieces by integrality of $b_i$. To see (b), observe that $x^*(P) \leq 2$, and thus, $y(P) \leq 1$, for every piece $P$ of auxiliary instance $\I'$. By feasibility of $x^*$, we have $x^*(P_i) \leq 1$, and thus, $y(P_i) \leq \frac{1}{2}$, for every commodity $(P_i,b_i)$ of the original instance $\I$ for which $b_i = 1$. \qedhere
\end{proof}

From now on, whenever we consider instances with $b = \one$, we may identify commodities $(P, 1)$ with their path $P$. The next lemma asserts that the neighborhood of the edge selected in iteration $t$ of the algorithm is sufficiently cheap. This will be essential when applying the fractional local ratio framework.

\begin{lemma}\label{lem:cheap_neighborhood_edge} 
Let $\I$ be an instance of $\BPPT$ defined on a tree $T = (V, E)$ rooted at $r \in V$ with $b = \one$. If $y\in [0,1]^{E}$ is a vector satisfying
$y(P)\le 1$ for every vertical commodity, and $y(P)\le \frac{1}{2}$ for every bent commodity, then there exists an edge $e\in E$ with
    $$
        y(N_{\I}[e]) \leq 1 + \frac{q_e}{4},
    $$
    where $q_e$ is the number of bent commodities that include edge $e$.
\end{lemma}

\begin{proof}
We write $N[e]$ rather than $N_{\I}[e]$ throughout this proof. For vertices $u, v \in V$ we denote by $P_{v,u}$ the unique $v$--$u$--path in $T$. If $u$ is a descendant of $v$, we say that $u$ is \emph{lower} than $v$ (and that $v$ is \emph{higher} than $u$). A vertex $v$ is called \emph{heavy} if there exists a vertex $u$ that is lower than $v$ such that $y(P_{v,u})\ge \tfrac{1}{4}$.
We select a leaf $\ell \in V$ as follows, and let $e \in E$ denote the unique edge incident to $\ell$.
\begin{itemize}
    \item If no vertex is heavy, $\ell$ is any leaf.
    \item Otherwise, let $z$ be a lowest heavy vertex, and let $\ell$ be a leaf such that $y(P_{\ell,z})\ge \tfrac{1}{4}$. 
\end{itemize}
By our choice of $e$, every commodity $P$ of $\I$ that contains $e$ has $\ell$ as an endpoint. Let $a_P$ denote the apex of such a path $P$, and let $v_P$ denote the endpoint of $P$ distinct from $\ell$ ($v_P = a_P$ in case $P$ is vertical). We decompose $P = P_{\ell,v_P}$ into the path $P_{\ell,a_P}$ which goes up from leaf $\ell$ to $a_P$ and the path $P_{a_P,v_P}$ which goes down from $a_P$ to $v_P$. 
If no path contains $e$, then $y(N[e]) = y_e \leq 1$ and we are done. Otherwise, let $a^*$ denote the highest apex $a_P$ of a commodity $P$ containing $e$. We can upper bound the $y$-mass of $N[e]$ as follows:
\begin{equation*}\label{eq:cost_neighborhood}
    y(N[e]) \leq y(P_{\ell,a^*}) + \sum_{\substack{P \text{ bent}\\ e \in E(P)}} y(P_{a_P,v_P})\,.
\end{equation*}
In order to prove the lemma, it is sufficient to bound each term separately. We distinguish the following two cases.
\begin{itemize}
    \item[a)] If no vertex is heavy, or if $a^*$ is lower than $z$, then $a^*$ is not heavy and no vertex that is lower than $a^*$ is heavy. This implies that the $y$-mass of $P_{\ell,a^*}$ is at most $1/4$, and also that the $y$-mass of every $P_{a_P,v_P}$ is at most $1/4$. Thus, $y(N[e])\le \frac{1}{4}+\frac{q_e}{4}$.
    
    \item[b)] Otherwise, $a^* = z$ or $z$ is lower than $a^*$. First, observe that $P_{\ell,a^*}$ is part of some path of $\I$, implying that its $y$-mass is at most $1$. Next, let $P$ be a bent commodity that contains $e$. Since $P$ is bent we have $y(P) \leq \frac12$. If $a_P$ is lower than $z$, then $a_P$ is not heavy and $y(P_{a_P,v_P}) \leq \tfrac{1}{4}$. Otherwise, $a_P$ lies on the path from $r$ to $z$. Since $y(P) \leq \frac{1}{2}$, we conclude that
    $
           y(P_{a_P,v_P}) = y(P) - y(P_{\ell,a_P}) \le y(P) - y(P_{\ell,z}) \leq \frac{1}{2} - \frac{1}{4} = \frac{1}{4}.
    $
    Again, we conclude that $y(N[e]) \le 1 + \frac{q_e}{4}$.\qed
\end{itemize}
\end{proof}

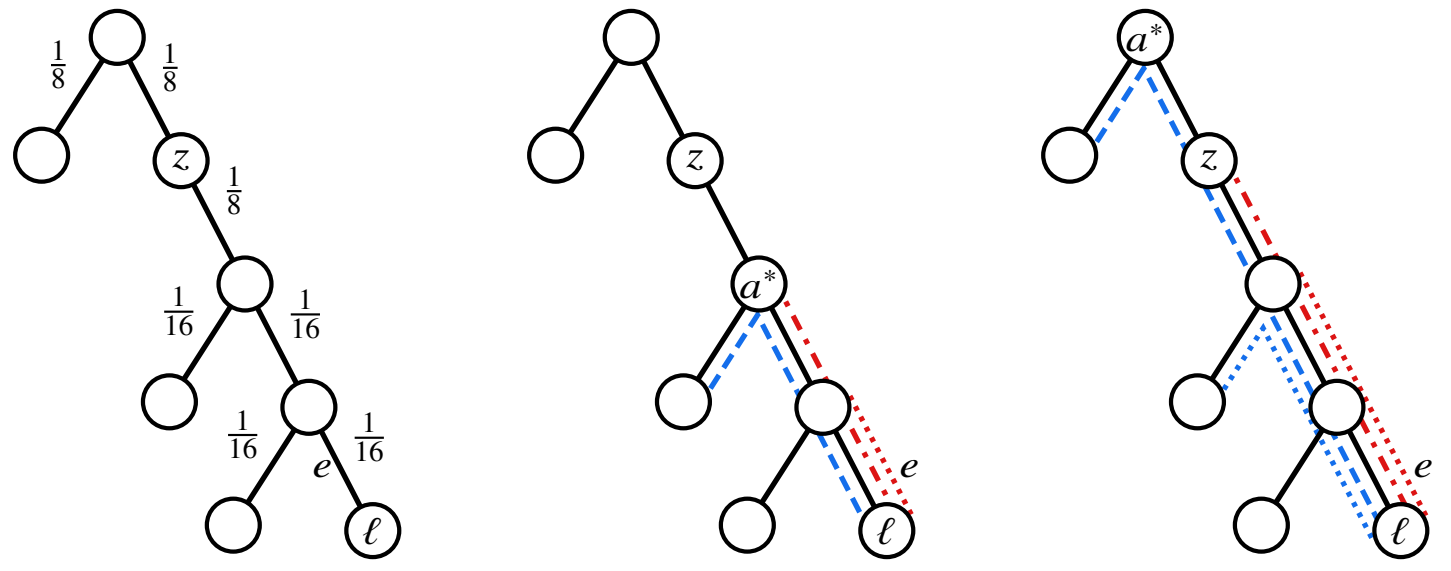
\begin{figure}[tbp]
    \centering
    \begin{tikzpicture}[yscale = 0.9,
    x=0.75cm, y=0.8cm,
    vtx/.style   = {circle, draw, thick, fill=white, minimum size=3.5mm, inner sep=0pt, font=\scriptsize},
    edge/.style  = {line width=1pt},
    lbl/.style   = {font=\scriptsize, inner sep=1.2pt},
    bentA/.style = {draw=cBlue, line width=1pt, dash pattern=on 3pt off 1.6pt},
    bentB/.style = {draw=cBlue, line width=1pt, dash pattern=on 0.9pt off 1.3pt},
    strA/.style  = {draw=cRed,  line width=1pt, dash dot},
    strB/.style  = {draw=cRed,  line width=1pt, dash pattern=on 0.9pt off 1.3pt},
  ]
  \definecolor{cBlue}{RGB}{20,110,235}
  \definecolor{cRed}{RGB}{220,20,20}

  \def\oA{3pt}\def\oB{5.6pt}

  \def\commodity#1#2#3{
    \foreach \d [count=\i] in {#2} {
      \expandafter\xdef\csname cmD\i\endcsname{\d}\xdef\cmDn{\i}
    }
    \foreach \n [count=\i from 0, remember=\n as \m (initially {})] in {#3} {
      \ifnum\i>0
        \ifnum\i>\cmDn
          \edef\cmOff{\csname cmD\cmDn\endcsname}
        \else
          \edef\cmOff{\csname cmD\i\endcsname}
        \fi
        \coordinate (cmS\i) at ($(\m)!\cmOff!90:(\n)$);
        \coordinate (cmE\i) at ($(\n)!\cmOff!-90:(\m)$);
      \fi
      \xdef\cmK{\i}
    }
    \coordinate (cmP0) at (cmS1);
    \foreach \i [evaluate=\i as \j using int(\i+1)] in {1,...,\cmK} {
      \ifnum\i<\cmK
        \pgfpointdiff{\pgfpointanchor{cmS\i}{center}}{\pgfpointanchor{cmE\i}{center}}
        \pgfgetlastxy{\cmAx}{\cmAy}
        \pgfpointdiff{\pgfpointanchor{cmS\j}{center}}{\pgfpointanchor{cmE\j}{center}}
        \pgfgetlastxy{\cmBx}{\cmBy}
        \pgfmathparse{abs(\cmAx*\cmBy-\cmAy*\cmBx)/(veclen(\cmAx,\cmAy)*veclen(\cmBx,\cmBy))}
        \ifdim\pgfmathresult pt<0.01pt
          \coordinate (cmP\i) at ($(cmE\i)!0.5!(cmS\j)$);
        \else
          \coordinate (cmP\i) at (intersection cs:
            first line={(cmS\i)--(cmE\i)}, second line={(cmS\j)--(cmE\j)});
        \fi
      \else
        \coordinate (cmP\i) at (cmE\i);
      \fi
    }
    \draw[#1] (cmP0) \foreach \i in {1,...,\cmK} { -- (cmP\i) };
  }

  \def\eL{1cm}\def\aS{-65}\def\aC{-120}
  \foreach \p/\dx in {A/0, B/3.4, C/6.8} {
    \path (\dx cm,0) coordinate (\p-r)
          ++(\aS:\eL) coordinate (\p-z)
          ++(\aS:\eL) coordinate (\p-v)
          ++(\aS:\eL) coordinate (\p-w)
          ++(\aS:\eL) coordinate (\p-l);
    \path (\p-r) ++(\aC:\eL) coordinate (\p-a);
    \path (\p-v) ++(\aC:\eL) coordinate (\p-b);
    \path (\p-w) ++(\aC:\eL) coordinate (\p-c);
  }

  \commodity{bentA}{-\oA}{B-b,B-v,B-w,B-l}
  \commodity{strA}{\oA}{B-v,B-w,B-l}
  \commodity{strB}{\oB}{B-w,B-l}

  \commodity{bentA}{-\oA}{C-a,C-r,C-z,C-v,C-w,C-l}
  \commodity{bentB}{-\oA,-\oB}{C-b,C-v,C-w,C-l}
  \commodity{strA}{\oA}{C-z,C-v,C-w,C-l}
  \commodity{strB}{\oB}{C-v,C-w,C-l}

  \foreach \p in {A,B,C} {
    \draw[edge] (\p-r) -- (\p-a)  (\p-r) -- (\p-z)  (\p-z) -- (\p-v)
                (\p-v) -- (\p-b)  (\p-v) -- (\p-w)
                (\p-w) -- (\p-c)  (\p-w) -- (\p-l);
    \foreach \x in {r,a,v,b,w,c} { \node[vtx] at (\p-\x) {}; }
    \node[vtx] at (\p-z) {$z$};
    \node[vtx] at (\p-l) {$\ell$};
  }

  \path (A-r) -- node[lbl, above left]  {$\tfrac{1}{8}$}  (A-a);
  \path (A-r) -- node[lbl, above right] {$\tfrac{1}{8}$}  (A-z);
  \path (A-z) -- node[lbl, above right] {$\tfrac{1}{8}$}  (A-v);
  \path (A-v) -- node[lbl, above left]  {$\tfrac{1}{16}$} (A-b);
  \path (A-v) -- node[lbl, above right] {$\tfrac{1}{16}$} (A-w);
  \path (A-w) -- node[lbl, above left]  {$\tfrac{1}{16}$} (A-c);
  \path (A-w) -- node[lbl, left]        {$e$}             (A-l);
  \path (A-w) -- node[lbl, above right] {$\tfrac{1}{16}$} (A-l);

  \node[font=\scriptsize, inner sep=0pt] at (B-v) {$a^{\ast}$};
  \path (B-w) -- node[lbl, right=7pt] {$e$} (B-l);

  \node[font=\scriptsize, inner sep=0pt] at (C-r) {$a^{\ast}$};
  \path (C-w) -- node[lbl, right=7pt] {$e$} (C-l);
\end{tikzpicture}
    \caption{Visualization of the case distinction in the proof of \Cref{lem:cheap_neighborhood_edge}. Numbers on the edges indicate $y$-values. Blue dotted and dashed commodities are bent unit-budget commodities and red dotted and dash-dotted commodities are vertical. The left graph shows the choice of the vertex $z$. The graph in the middle shows case a) in the case distinction, where $a^*$ is a proper descendant of $z$. Lastly, the graph on the right shows case b) where $a^*$ is an ancestor of $z$ (or $a^* = z$).
    \label{fig:neighborhood_lemma}}
\end{figure}
A visualization of the case distinction in the proof of \Cref{lem:cheap_neighborhood_edge} can be seen in \Cref{fig:neighborhood_lemma}.
The proof of \Cref{thm:two_plus_uc_half_approx} is a standard application of the fractional local ratio framework. Thus, we defer the full proof to Appendix \ref{appendix:proof_two_uc_half_apx} and present just a sketch here. \medskip

\noindent\textit{Sketch of the proof of \Cref{thm:two_plus_uc_half_approx}.}
By \Cref{lem:properties_i_prime}, the original auxiliary instance $\I'_0 \coloneqq \I'$ fulfills the requirements of \Cref{lem:cheap_neighborhood_edge}, and this also holds for the instances $\I'_t$ constructed during the run of the algorithm by contracting edges in $\I'_{t-1}$. The core idea behind the proof of the approximation guarantee is the observation that the weight $w_t(e_t)$ subtracted in iteration $t$ is counted at least once by the final solution $F_{\ALG}$ which contains $e_t$ or one of its neighbors. Meanwhile, the scaled LP solution $y = x^*/2$ counts the same weight at most $y(N_{\I'_t}[e_t]) \leq 1 + \frac{\ucong}{4}$ times by \Cref{lem:cheap_neighborhood_edge} and $q_e \leq \ucong$.
Summing over all iterations, we obtain $w(F_{\ALG}) \geq w^\intercal y/(1+\frac{UC}{4}) = \OPT_{\text{LP}}/(2+\frac{\UC}{2})$. \medskip

\noindent\textbf{Parameter $\conge$.} In Appendix~\ref{app:flr_parameter_congestion_improved}, we apply a similar approach to the one of \Cref{thm:two_plus_uc_half_approx} to prove an approximation guarantee for $\BPPT$ parameterized in $\conge$ rather than $\ucong$. The new approach differs from \Cref{alg:2_plus_uc_half} only in the construction of the auxiliary instance and the proof of the small local neighborhood lemma; the construction of the solution via fractional local ratio remains unchanged. The key new insight is that the auxiliary instance only needs to satisfy two properties: (a) commodities are split into at most $b_i$ pieces to guarantee feasibility, and (b) the auxiliary instance always contains an edge with a sufficiently small LP mass in its neighborhood. In particular it is not necessary that the constraints in the auxiliary instance are given by paths, and in fact the constructed instance contains commodities consisting of two disjoint path segments. With this adapted approach, we prove the following theorem.

\begin{restatable}{theorem}{flrCongestionImproved}
\label{thm:flrCongestionImproved}
    There exists a polynomial-time algorithm that, given a \BPPT{} instance, computes a feasible solution of value at least $\OPT_{\text{LP}}/(1+\conge/2)$.
\end{restatable}

\subsection{Integrality Gap}\label{sec:two_uc_half_integrality_gap}

Closing this section, we sketch the construction used to prove \Cref{thm:two_plus_uc_half_integrality_gap} which states that there exists a family of instances of $\BPPT$ for which the integrality gap of the LP converges to $2+\ucong/2$, implying that the approximation guarantee obtained by the algorithm described above is tight. In this subsection, we show how to construct the instance and present a feasible LP solution $x$. The proofs for showing that $x$ is indeed feasible, and that there exists an upper bound on the weight of any integral solution for the instance which is at least a multiplicative factor of $2+\frac{\UC}{2}$ away from the weight of the fractional solution $x$ are deferred to Appendix \ref{appendix:two_plus_uc_half_integrality_gap}.\medskip

\noindent\textbf{Construction of the instance.} We fix some $\UC \geq 0$ and $n \geq 2$, and recursively define a family of rooted trees $T_d$ for $d\in \{1,2, \ldots \}$ as follows: $T_1$ is a star with $\UC + 1$ leaves rooted at its center. To construct $T_d$ for $d\ge 2$, take a root $r$ and attach $n$ copies $T_{d-1}^{(1)},\ldots,T_{d-1}^{(n)}$. For $i\in[n]$, connect $r$ to the root of the $i$-th copy $T_{d-1}^{(i)}$ with an edge $e_i$.

We say that a \emph{maximal chain} of $T_d$ is a root-to-leaf path of $T_d$ and denote the set of all maximal chains of $T_d$ by $\mathcal{Q}_d$. Similarly, $\mathcal{Q}_{d-1}^{(i)}$ denotes the set of all maximal chains in the $i$-th copy of $T_{d-1}$ used in the construction of $T_d$. We refer to the edges incident to leaves (i.e., corresponding to edges in stars) as \emph{pendant edges} and define the edge weights $w_e$ of pendant edges as $\UC + 1$. For every other edge $e$, we define $w_e$ to be the number of leaves below edge $e$ (when directing all edges away from the root $r$). That is, for an edge $e$ we have $w_e \coloneqq \UC + 1$ if $e$ is a pendant edge, and $w_e \coloneqq|\{Q \in \mathcal{Q}_d \ | \ e \in Q\}|$ otherwise.
Using these weights, the total weight of every level except the lowest of $T_d$ is exactly the number of leaves in $T_d$, that is, $(\UC + 1) n^{d-1}$. The last level consequently has weight $(\UC + 1) \cdot (\UC + 1) n^{d-1}$.

Next, we define for every $d$ two sets of commodities, called $\C_d$ and $\U_d$. We set $\C_1 = \emptyset$ and for $d > 1$, the set $\C_d$ consists of the following commodities, with $b \equiv 2$: The union of all commodities $\C_{d-1}^{(i)}$ over all $i \in [n]$ and the new commodities $\{\{e_i, e_j\} \cup Q \ | \ i \neq j, \ Q \in \mathcal{Q}_{d-1}^{(i)}\}$ consisting of a maximal chain of $T_d$ plus one further edge adjacent to $r$.

\begin{figure}[tbp]
    \centering
    \begin{tikzpicture}[yscale=0.8, xscale = 0.9]
  \definecolor{tBlue}{RGB}{20,110,235}
  \definecolor{tRed}{RGB}{220,20,20}
  \tikzset{
    t3 vtx/.style   = {circle, draw, fill=white, line width=0.6pt, inner sep=0pt, minimum size=8.0pt},
    t3 leaf/.style  = {circle, draw, fill=white, line width=0.6pt, inner sep=0pt, minimum size=6pt},
    t3 edge/.style  = {thick},
    t3 lbl/.style   = {font=\small},
    t3 brace/.style = {decorate, decoration={brace, amplitude=4pt}, thick},
    t3 blue1/.style = {draw=tBlue, line width=0.8pt, dash pattern=on 3pt off 1.6pt},
    t3 blue2/.style = {draw=tBlue, line width=0.8pt, dash dot},
    t3 red1/.style  = {draw=tRed,  line width=0.8pt, dash pattern=on 0.9pt off 1.1pt},
    t3 red2/.style  = {draw=tRed,  line width=0.8pt, dash pattern=on 3pt off 1.6pt},
  }

  \def\tCommodity#1#2#3{%
    \foreach \td [count=\ti] in {#2} {%
      \expandafter\xdef\csname tD\ti\endcsname{\td}\xdef\tDn{\ti}%
    }%
    \foreach \tn [count=\ti from 0, remember=\tn as \tm (initially {})] in {#3} {%
      \ifnum\ti>0
        \ifnum\ti>\tDn
          \edef\tOff{\csname tD\tDn\endcsname}%
        \else
          \edef\tOff{\csname tD\ti\endcsname}%
        \fi
        \expandafter\xdef\csname tO\ti\endcsname{\tOff}%
        \coordinate (tS\ti) at ($(\tm)!\tOff!90:(\tn)$);
        \coordinate (tE\ti) at ($(\tn)!\tOff!-90:(\tm)$);
      \fi
      \xdef\tK{\ti}%
    }%
    \coordinate (tP0) at (tS1);
    \foreach \ti [evaluate=\ti as \tj using int(\ti+1)] in {1,...,\tK} {%
      \ifnum\ti<\tK
        \pgfmathanglebetweenpoints{\pgfpointanchor{tS\ti}{center}}{\pgfpointanchor{tE\ti}{center}}%
        \let\tAa\pgfmathresult
        \pgfmathanglebetweenpoints{\pgfpointanchor{tS\tj}{center}}{\pgfpointanchor{tE\tj}{center}}%
        \let\tAb\pgfmathresult
        \pgfmathtruncatemacro\tBevel{((\csname tO\ti\endcsname)*(\csname tO\tj\endcsname) < 0)
            || (abs(sin(\tAb-\tAa)) < 0.02)
            || ((cos(\tAb-\tAa) < 0) && ((\csname tO\ti\endcsname)*sin(\tAb-\tAa) < 0))}%
        \ifnum\tBevel=1
          \coordinate (tA\ti) at (tE\ti);
          \coordinate (tB\ti) at (tS\tj);
        \else
          \coordinate (tA\ti) at (intersection cs:
            first line={(tS\ti)--(tE\ti)}, second line={(tS\tj)--(tE\tj)});
          \coordinate (tB\ti) at (tA\ti);
        \fi
      \else
        \coordinate (tA\ti) at (tE\ti);
        \coordinate (tB\ti) at (tE\ti);
      \fi
    }%
    \draw[#1] (tP0) \foreach \ti in {1,...,\tK} { -- (tA\ti) -- (tB\ti) };
  }

  \coordinate (r) at (0,0);
  \foreach \i/\xi in {1/-2.5, 2/2.5} {
    \coordinate (v\i) at (\xi,-1.1);
    \foreach \j/\dj in {1/-1.05, 2/1.05} {
      \coordinate (v\i\j) at ({\xi+\dj},-2.2);
      \foreach \k/\dk in {1/-0.51, 2/-0.17, 3/0.17, 4/0.51} {
        \coordinate (v\i\j\k) at ({\xi+\dj+\dk},-3.3);
      }
    }
  }

  \tCommodity{t3 blue1}{1.8pt}{v111,v11,v1,r,v2}
  \tCommodity{t3 blue2}{-1.8pt}{v114,v11,v1,v12}
  \tCommodity{t3 red1}{1.8pt}{v121,v12,v124}
  \foreach \x/\sa/\sb in {v122/12.3pt/29.1pt, v123/11.75pt/28.55pt} {
    \coordinate (a1) at ($(v12)!\sa!(\x)$);
    \coordinate (a2) at ($(v12)!\sb!(\x)$);
    \draw[t3 red2] ($(a1)!1.8pt!-90:(\x)$) -- ($(a2)!1.8pt!-90:(\x)$);
  }

  \foreach \i in {1,2} {
    \draw[t3 edge] (r) -- (v\i);
    \foreach \j in {1,2} {
      \draw[t3 edge] (v\i) -- (v\i\j);
      \foreach \k in {1,2,3,4} { \draw[t3 edge] (v\i\j) -- (v\i\j\k); }
    }
  }
  \node[t3 vtx, label={[font=\small]above:$r$}] at (r) {};
  \foreach \i in {1,2} {
    \node[t3 vtx] at (v\i) {};
    \foreach \j in {1,2} {
      \node[t3 vtx] at (v\i\j) {};
      \foreach \k in {1,2,3,4} { \node[t3 leaf] at (v\i\j\k) {}; }
    }
  }

  \draw[t3 brace] ([xshift=0.07cm, yshift=-0.17cm]v114) -- node[below=3pt, xshift=0.25cm, inner sep=1pt, font=\footnotesize] {$(UC+1)$ children}
                  ([xshift=-0.07cm, yshift=-0.17cm]v111);
  \node[t3 lbl, anchor=west] at (1.4,-0.3) {$n$ children};

  \foreach \yt/\yb/\dep/\wt/\xv in {0/-1.1/1/8/\tfrac{3}{8}, -1.1/-2.2/2/4/\tfrac{3}{4}, -2.2/-3.3/3/4/\tfrac{1}{2}} {
    \pgfmathsetmacro\ym{(\yt+\yb)/2-0.1}
    \draw[t3 brace] (4.45,\yt-0.05) -- (4.45,\yb+0.05);
    \node[t3 lbl, anchor=base west] at (4.7,\ym) {depth $\dep$};
    \node[t3 lbl, anchor=base west] at (6.2,\ym+0.2) {$w(e) = \wt$};
    \node[t3 lbl, anchor=base west] at (6.2,\ym-0.24) {$x(e) = \xv$};
  }
\end{tikzpicture}
    \caption{The construction used in the proof of \Cref{thm:two_plus_uc_half_integrality_gap} showing $T_3$ with $\UC = 3$ and $n = 2$. Dashed and dash-dotted blue edges are commodities in $\C$, and dashed and dotted red edges correspond to commodities in $\U$. Only some of the commodities are shown.}\label{fig:two_plus_uc_half_integrality_gap}
\end{figure}
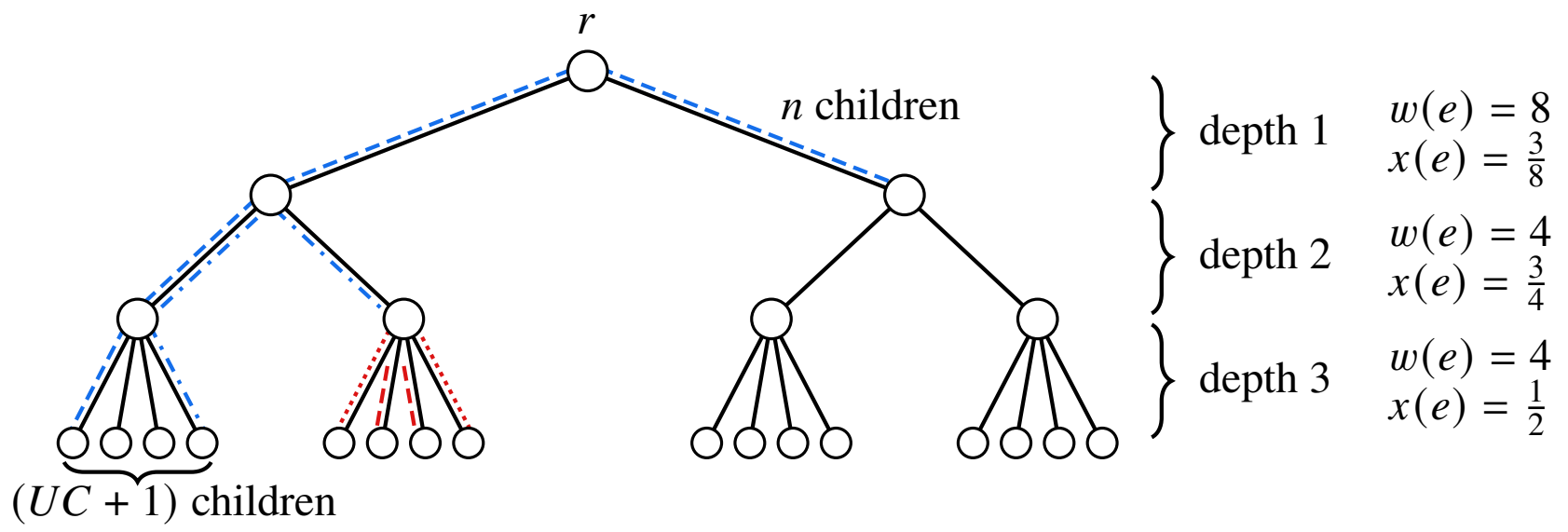

For $\U$, we set $\U_1 = \{ \{f, f' \} \mid f \text{ and } f' \text{ are edges in } T_1 \text{ with } f \neq f'\}$ as the set of all pairs of edges of $T_1$, with $b \equiv 1$. For $d > 1$, we define $\U_d = \bigcup_{i \in [n]} \U_{d-1} ^{(i)}$ as the union of all the $\U$ commodities that are part of one of the copies $T_{d-1}^{(i)}$. A visualization of the construction and the fractional LP solution $x$ can be seen in \Cref{fig:two_plus_uc_half_integrality_gap}.
The proof of \Cref{thm:two_plus_uc_half_integrality_gap} follows readily from the bounds $\OPT \le (\UC + 1) n^{d-1}(1 + \frac{1}{n})$ and $\OPT_{LP} \ge (\UC + 1) n^{d-1} \big(\frac{\UC + 1}{2} + 3 (2^{-1} - 2^{-d})\big)$, see Appendix~\ref{appendix:two_plus_uc_half_integrality_gap} for details.

\section{A $\tfrac{\mincap}{\mincap - 1}$-approximation Algorithm}
\label{sec:lp_guided}

In this section, we present an LP-guided splitting algorithm yielding an approximation guarantee of $\tfrac{\mincap}{\mincap - 1}$ for $\mincap \geq 2$. As a corollary, we obtain weight at least $\OPT_{\LP}$ if we are allowed to violate the budget constraints by at most one for each commodity. We also note that the $\tfrac{\mincap}{\mincap - 1}$ guarantee coincides with the integrality gap of \eqref{eq:packing_LP} relative to $\beta$, the proof of which can be found in Appendix~\ref{appendix:ell_ell_minus_one_integrality_gap}. We remark that $\BPPT$ remains $\NP$-hard for every fixed $\beta \geq 2$. This can easily be seen by extending the construction from the reduction from MWIS to a subdivided star with outer edges of sufficiently high weight.

\begin{restatable}{theorem}{EllEllMinusOneApx}
\label{thm:ell_ell_minus_one_approx}
    There is a polynomial-time algorithm that, given a \BPPT{} instance with $\mincap \geq 2$, computes a feasible solution of value at least $\tfrac{\mincap - 1}{\mincap}\OPT_{LP}$.
\end{restatable} 

\vspace{-1ex} 

\begin{restatable}{theorem}{EllEllMinusOneIntegralityGap}
\label{thm:ell_ell_minus_one_integrality_gap}
    For every fixed $\mincap \geq 2$, the integrality gap of the standard LP relaxation of \BPPT{} is exactly $\tfrac{\mincap}{\mincap-1}$.
\end{restatable}

\paragraph{Description of the algorithm.} Given an instance $\I$ with $\beta\ge 2$, our algorithm starts with an optimal solution $x^*$ to \eqref{eq:packing_LP} and chooses an arbitrary root $r$ in the tree. Assume that $\I$ contains at least one bent path, otherwise we can solve the instance exactly by total unimodularity. We split each bent path $P_i$ into its two sides $L_i$ and $R_i$ at its apex, assigning the two new paths a budget of $\lceil x^*(L_i)/\alpha \rceil$ and $\lceil x^*(R_i)/\alpha \rceil$, respectively, where $\alpha := \max\left\{\frac{b_j}{b_j - 1} \mid P_j \text{ bent}\right\}$. Lastly, the algorithm returns an optimal integral solution $F_{\ALG}$ for the resulting vertical instance $\I'$, this is possible due to total unimodularity. The pseudocode of the algorithm can be found in Appendix~\ref{appendix:ell_ell_minus_one_pseudocode}.

\begin{proof}[of \Cref{thm:ell_ell_minus_one_approx}]
    We claim that the algorithm described  above attains the desired approximation guarantee. First, observe that $x^*/\alpha$ is feasible for the auxiliary instance $\I'$ constructed by the algorithm and that $\alpha \leq \frac{\mincap}{\mincap - 1}$. Hence, it suffices to show that the integral solution $F_{\ALG}$ as constructed in the algorithm is feasible for the original instance $\I$ to obtain the desired approximation guarantee. Note that vertical commodities in $\I$ remain vertical in the auxiliary vertical instance $\I'$. Hence, we have $|F_{\mathrm{ALG}}\cap P_i|\le b_i$ for every vertical path $P_i$. For every bent commodity $(P_i,b_i)$, observe that
    $\alpha\ge\frac{b_i}{b_i-1}$ and $x^*(P_i)\le b_i$, implying $x^*(P_i)/\alpha\le b_i-1$. Thus,
    \[ 
        |F_{\mathrm{ALG}}\cap P_i| =
        |F_{\ALG} \cap L_i| + |F_{\ALG} \cap R_i| \leq \left\lceil\frac{x^*(L_i)}{\alpha}\right\rceil+\left\lceil\frac{x^*(R_i)}{\alpha}\right\rceil
      < \frac{x^*(P_i)}{\alpha}+2 \leq 
        b_i+1, 
    \]
    so $|F_{\mathrm{ALG}}\cap P_i|\le b_i$ by integrality. \qed
\end{proof}

We defer the proof of the integrality gap tightness of our guarantee to Appendix~\ref{appendix:ell_ell_minus_one_integrality_gap}.
Lastly, observe that by slightly adjusting the budget that we assign each side in the above algorithm, we can obtain the following result.

\begin{corollary}
    For arbitrary instances of $\BPPT$, we can efficiently compute a relaxed solution $F_{\ALG}$ of weight at least $\OPT_{\text{LP}}$ satisfying $|F_{\ALG} \cap P_i| \leq b_i + 1$ for all $i \in [m]$.
\end{corollary}
\begin{proof}
    By adjusting the budgets of the new commodities in the above algorithm  to be $ \lceil x^*(L_i) \rceil$ and $ \lceil x^*(R_i) \rceil$, respectively, and denoting the computed solution by $F_{\ALG}$, we observe the that $|F_{\ALG} \cap P_i| \leq \lceil x^*(L_i) \rceil + \lceil x^*(R_i)\rceil < x^*(L_i) + x^*(R_i) + 2 \leq b_i + 2$.
    Hence, $|F_{\ALG} \cap P_i| \leq b_i + 1$ by integrality and $F_{\ALG}$ satisfies the desired feasibility constraint. Furthermore, by construction, $F_{\ALG}$ has weight at least $\OPT_{\LP}$. \qed
\end{proof}
\ifsubmission
\else
\section*{Acknowledgements}
Samuel Fiorini acknowledges funding from \emph{King Baudouin Foundation} through project BD-DELTA-2 (convention 2023-F2150080-233051, 2023--26) and \emph{Fonds de la Recherche Scientifique} - FNRS through research project BD-DELTA-3 (PDR 40028812, 2024--27).
Philipp Pabst acknowledges funding from the \emph{DFG} through the Research Training Group UnRAVeL (project number
282652900).
We thank Lennart Kauther and Khai Van Tran for fruitful discussions.
\fi

\section*{AI Acknowledgment}
We obtained several initial results on $\BPPT$ without the help of AI and used Claude Opus 5.5 to further improve them, leading to the results presented in this paper.

For the $(2 + \ucong/2)$-approximation presented in \Cref{sec:2_plus_uc_half}, we presented the model with a $2(\ucong + 1)$-approximation using an approach similar to the splitting procedure from \Cref{sec:lp_guided} with an additional repair step to deal with violated budgets for unit-budget commodities. Claude Opus 5.5 then used the fractional local ratio framework to first obtain a $(\conge + 1)/2$-approximation restricted to instances with $b = \one$. Later, the model found the constructions of the auxiliary instances $\I'$ used to prove \Cref{thm:two_plus_uc_half_approx,thm:flrCongestionImproved} and the construction to show integrality gap tightness presented in the proofs of \Cref{thm:two_plus_uc_half_integrality_gap,thm:ell_ell_minus_one_integrality_gap}. To prove \Cref{thm:congestion_approx_halldorsson_lau}, we asked the model to generalize the result of Halldórsson and Lau using a notion of conflict graphs at every vertex similar to the definition of the link graphs $H_v$ from Appendix \ref{sec:coloring_approx}. Claude Opus 5.5 then managed to extend the approach, first for $b = \one$ and later for general budgets.

The authors verified, rewrote and simplified the results generated by AI and take full responsibility for the correctness of all results presented in this paper.

\ifsubmission
\else
\newpage
\fi

\bibliographystyle{splncs04}
\bibliography{references}

\newpage

\appendix

\renewcommand{\theHsection}{\Alph{section}}

\section{Hypergraph $b$-matching is a $k$-exchange System}
\label{app:kexchange}

In this section we show that the set of $b$-matchings in rank-$k$ hypergraphs is a $k$-exchange system in the sense of Feldman, Naor, Schwartz and Ward \cite{feldman11kexchange}. This immediately implies a $(\frac{k+1}{2}+\varepsilon)$-approximation by non-oblivious local search in $\XP$-running time parameterized in $k$. By naively analyzing the running time, one obtains a bound of $O(\varepsilon^{-2} \cdot |E|^{O(k^2)})$. Furthermore, since every $k$-exchange system is a $k$-extendible system in the sense of Mestre \cite{mestre06extendible}, the greedy algorithm computes a $k$-approximation in fully polynomial running time.

As noted in the introduction, $\BPPT$ is a special case of weighted $b$-matching in rank-$\conge$ hypergraphs. Thus, results for $k$-exchange systems immediately yield approximation results for $\BPPT$ (even on general graphs rather than trees). As a start, recall that a tuple $(E, \mathcal{F})$ consisting of a finite set $E$ and a collection of subsets $\mathcal{F}\subseteq 2^E$ is called an \emph{independence system} if $\mathcal{F}$ is downward-closed. $k$-exchange systems form a special class of independence systems, and  are defined as follows:

\begin{definition}[see \cite{feldman11kexchange}]
    An independence system $(E, \mathcal{F})$ is an $k$-exchange system if for all $A,B \in \mathcal{F}$ there exists a family $\{Y_e \subseteq B \setminus A \ | \ e \in A \setminus B\}$ such that the following conditions hold:
    \begin{enumerate}
        \item[(K1)] $|Y_e| \leq k$ for all $e \in A \setminus B$,
        \item[(K2)] every $f \in B \setminus A$ belongs to at most $k$ sets $Y_e$,
        \item[(K3)] for all $A' \subseteq A \setminus B$ it holds that
        \[\Big(B \setminus \bigcup_{e \in A'} Y_e\Big) \cup A' \in \mathcal{F}.\]
    \end{enumerate}
\end{definition}

We show that the set of feasible $b$-matchings in a rank-$k$ hypergraph indeed forms a $k$-exchange system. As a consequence, given an instance of weighted $b$-matching in rank-$k$ hypergraphs, we can compute a feasible solution $F$ satisfying $w(F) \geq \OPT/(\tfrac{k+1}{2} + \varepsilon)$ using non-oblivious local search.
\begin{observation}
\label{thm:kexchange}
    Let $H = (V,E)$ be a rank-$k$ hypergraph, that is, $|e| \leq k$ for all hyperedges $e \in E$, with budgets $b: V \rightarrow \Z_{\geq 1}$. Then, the set
    \[\mathcal{F} \coloneqq \{F \subseteq E : |\{e \in E \ | \ e \in F, v \in e\}| \leq b_v \ \forall v \in V\}\] is a $k$-exchange system.
\end{observation}

\begin{proof}
Clearly $\mathcal F$ is an independence system. Let $A,B\in\mathcal F$ and,
for $v\in V$, define $\delta(v):=\{e\in E: v\in e\}$. For each $v\in V$, fix a
maximum matching $M_v$ in the complete bipartite graph with sides
$\delta(v)\cap(B\setminus A)$ and $\delta(v)\cap(A\setminus B)$, and write
$M_v(e)$ for the partner of a matched element~$e$. For $e\in A\setminus B$ set
\[ Y_e:=\{M_v(e) : v\in e,\ e \text{ matched in } M_v\}\subseteq B\setminus A. \]
\textbf{(K1)} We have $|Y_e|\le |e|\le k$.

\noindent\textbf{(K2)} If $f\in Y_e$, then $e=M_v(f)$ for some $v\in f$. As each $M_v$
is a matching, $f$ has at most one partner per $v\in f$, so $f$ lies in at most
$|f|\le k$ sets~$Y_e$.

\noindent\textbf{(K3)} Let $A'\subseteq A\setminus B$, $B':=(B\setminus\bigcup_{e\in A'}Y_e)\cup A'$
and $v\in V$. The partners $M_v(e)$ of the matched $e\in A'\cap\delta(v)$ are
distinct elements of $B\cap\delta(v)$ and are removed, so
\[ |B'\cap\delta(v)| \le |B\cap\delta(v)| + |\{e\in A'\cap\delta(v): e \text{ unmatched in } M_v\}|. \]
If no element of $A'\cap\delta(v)$ is unmatched, this is at most $b_v$.
Otherwise $M_v$ covers all of $\delta(v)\cap(B\setminus A)$ by maximality, so
the number of unmatched elements is at most
$|(A\setminus B)\cap\delta(v)|-|(B\setminus A)\cap\delta(v)|
 =|A\cap\delta(v)|-|B\cap\delta(v)|$, giving $|B'\cap\delta(v)|\le|A\cap\delta(v)|\le b_v$. \qed
\end{proof}

\section{Missing Proofs for the $(2 + \ucong/2)$-approximation}

\subsection{Pseudocode}\label{appendix:2_UC_half_pseudocode}

Pseudocode of the algorithm attaining a $2 + \frac{\UC}{2}$-approximation is shown in \Cref{alg:2_plus_uc_half}.

\begin{algorithm}
    \caption{Fractional Local Ratio}
    \label{alg:2_plus_uc_half}
    
    \textbf{Input:} A $\BPPT$ instance consisting of a tree $T$, commodities $(P_i, b_i)_{i \in [m]}$ and weights $(w_e)_{e\in E}$
    \textbf{Output:} A feasible solution $F_{\ALG}$
    \begin{algorithmic}[1]
    
    \vspace*{0.2cm}
    \Statex \emph{Building the auxiliary instance:}
    \State Compute an optimal fractional solution $x^*$ to (LP)
    \State Choose an arbitrary root $r \in V$
    \State Split all bent commodities $P_i$ with $b_i \geq 2$ into commodities $(L_i, b_i)$ and $(R_i, b_i)$ at their apex.
    \For{commodity $(P_j, b_j)$ with $b_j \geq 2$}
        \For{both sides of $P_j$}
            \parState{Starting at the apex, walk along the path away from the root. Sum up the $x^*$-values of the edges on the path.}
            \parState{Whenever the value of the sum would exceed $2$ when adding the $x^*$-value of the next edge, split the current path into two at the common vertex of the current and next edge.}
            \State Repeat this with the remaining path until the end of the path is reached.
        \EndFor
    \EndFor
    \State Denote the constructed auxiliary instance by $\I'$.
    
    \vspace*{0.2cm}
    \Statex \emph{Choose edges based on fractional local ratio:}
    \State Initialize $y \coloneqq x^* / 2, t \coloneqq 0, w_0 \coloneqq w$, and $\I'_0 \coloneqq \I'$.
    \While {there exists an edge $e$ in $\I'_t$}
        \State Find $e$ minimizing $y(N_{\I'_t}[e])$ and set $e_t = e$.
        \State Set $w_{t+1}(f) = w_t(f) - w_t(e_t)$ for all $f \in N_{\I'_t}[e_t]$ and $w_{t+1}(f) = w_{t}(f)$ otherwise.
        \State Let $\I'_{t+1}$ be the instance obtained by contracting all edges $e$ in $\I'_t$ for which $w_{t+1}(e) \leq 0$.
        \State Set $t = t+1$.
    \EndWhile

    \vspace*{0.2cm}
    \Statex \emph{Greedily choose edges in reverse:}
    \State $F_{\ALG} = \emptyset$.
    \For{$e \in \{e_{t-1}, \ldots, e_0\}$}
        \If {$F_{\ALG} \cup \{e\}$ is feasible in $\I'$}
            \State Add $e$ to $F_{\ALG}$.
        \EndIf
    \EndFor
    
    \State \Return $F_{\ALG}$
    \end{algorithmic}
\end{algorithm}

\subsection{Proof of \Cref{thm:two_plus_uc_half_approx}}\label{appendix:proof_two_uc_half_apx}

\begin{proof}
We claim that the algorithm described in \Cref{sec:2_plus_uc_half_algo_descr} computes a $(2+\ucong/2)$-approximation. First, note that the algorithm runs in polynomial time, terminates and outputs an integral solution $F_{ALG}$. The computed solution is feasible for $\I'$, and thus also feasible for $\I$ by \Cref{lem:properties_i_prime}.

We bound the cost $w(F_{\ALG})$ of our computed solution  by bounding it with respect to each iteration $t$, as usual in fractional local ratio. To this end, define $\delta_t$ as the decrement for iteration $t$. That is, $\delta_t(f) = \min\{w_t(e_t), w_t(f)\}$ for $f \in N_{\I'_t}[e_t]$, and $\delta_t(f) = 0$ otherwise. Now, we have $w = \sum_t \delta_t$ and in particular $w(F_{\ALG}) = \sum_t \delta_t(F_{\ALG})$.  

To bound the cost of iteration $t$, observe that \Cref{lem:cheap_neighborhood_edge} is applicable since its requirements hold for $\I'_0 = \I'$ by \Cref{lem:properties_i_prime} and for later iterations since contraction keeps vertical commodities vertical and does not increase the $y$-mass of commodities. Thus, there exists an edge $e \in E(\I'_t)$ such that $y(N_{\I'_t}[e]) \leq 1 + \frac{q_e}{4} \leq 1 + \frac{\UC(\I)}{4}$. The latter holds since the bent commodities in $\I'_t$ correspond to unit-budget commodities in $\I$.
Now, since the algorithm chooses an $e$ that minimizes $y(N_{\I'_t}[e])$, we obtain $y(N_{\I'_t}[e_t]) \leq 1 + \frac{\UC}{4}$.
With this, we obtain the following locality argument
\begin{equation}\label{eq:local_lower_bound_on_decrement}
    \delta_t^\intercal y = \sum_{f \in E(\I'_t)}\delta_t (f) y(f) \leq \delta_t(e_t) y(N_{\I'_t}[e_t]) \leq \left(1 + \frac{\UC}{4}\right) \delta_t(e_t).
\end{equation}

Lastly, we note that $\delta_t(F_{\ALG}) \geq \delta_t (e_t)$. This is clear if $e_t \in F_{\ALG}$. Otherwise, since every commodity has budget 1 in $\I'$, there exists an $e_s \in F_{\ALG}$ with $s>t$ such that $e_s \in N_{\I'_t}[e_t]$. Now, since $s > t$, we have $w_t(e_s) \geq w_t(e_t)$ and thus $\delta_t(F_{\ALG}) \geq \delta_t (e_t)$. Combining this with (\ref{eq:local_lower_bound_on_decrement}) and summing over all iterations, we obtain
\begin{multline*}
    w(F_{\ALG}) = \sum_t \delta_t(F_{\ALG}) 
    \geq \sum_t \delta_t(e_t) \\
    \overset{(\ref{eq:local_lower_bound_on_decrement})}{\geq} \frac{\sum_t \delta_t ^\intercal y}{1 + \UC/4} =
    \frac{w^\intercal y}{1 + \UC/4} =
    \frac{w^\intercal x^*}{2 + \UC/2} \geq
    \frac{\OPT_{\text{LP}}}{2 + \UC/2}.\qedhere
\end{multline*}
\end{proof}

\subsection{Integrality Gap}\label{appendix:two_plus_uc_half_integrality_gap}
Recall the construction of $T_d$, $\C$ and $\U$ as described in \Cref{sec:two_uc_half_integrality_gap}. To prove that the integrality gap of the LP is $2+\frac{\UC}{2}$ (i.e., \Cref{thm:two_plus_uc_half_integrality_gap}), it remains to prove the following two results.

\begin{restatable}{lemma}{l.integralitygap1}
\label{lemma:l.integralitygap1}
The weight of any integral solution for \BPPT\ on $T_d$ can be upper bounded by $(\UC + 1) n^{d-1}(1 + \frac{1}{n})$.
\end{restatable}

\vspace{-3ex}

\begin{restatable}{lemma}{l.integralitygap2}
\label{lemma:l.integralitygap2}
Given instance $T_d$ as constructed above, set $x(e) = \frac{1}{2}$ for every pendant edge, and
    $x(e) = 3 \cdot 2^{j-d-1}$ for every depth-$j$ edge of $T_d$ for $j < d$.
Then $x$ is a feasible LP-solution of
 weight at least $(\UC + 1) n^{d-1} (\frac{\UC + 1}{2} + 3 (2^{-1} - 2^{-d})).$
\end{restatable}

For the proof of \Cref{lemma:l.integralitygap1}, the following claim turns out to be helpful.

\begin{claim}[1]\label{claim1} 
    Let $F'$ be any integral solution of $\BPPT$ on $T_d$  such that $F' \cap T_{d-1}^{(i)}$ intersects each maximal chain $Q \in \mathcal{Q}_{d-1}^{(i)}$ of the subtree $T_{d-1}^{(i)}$ at most once. Then  $w(F' \cap T_{d-1}^{(i)}) \leq (\UC + 1) n^{d-2}$.
\end{claim}
\begin{proof}
We assume $d > 2$, since the case $d \leq 2$ is trivial. Given a solution $F'$ satisfying the property above, we push all its edges in $T_{d-1}^{(i)}$ down to the second to lowest level: We replace each edge $e \in F' \cap T_{d-1}^{(i)}$ above depth $d-1$ by all depth $d-1$ edges below it, that is, those lying in a maximal chain $Q\in\mathcal{Q}_{d-1}^{(i)}$ with $e$. In this process, (i) the number of cuts on each maximal chain stays constant, (ii) the weight of the solution stays constant, since $w(e)$ is the number of leaves under $e$, and (iii) the solution remains feasible. 
Similarly, we push pendant edges one level up while remaining feasible with the same weight to obtain a solution with only edges of depth $d-1$.
Recall that the weight of every level except the last level in $T_{d-1}^{(i)}$ is exactly $(\UC + 1) n^{d-2}$. Thus, the weight of $F' \cap T_{d-1}^{(i)}$ is at most the weight of level $d-1$ of $T_{d-1}^{(i)}$ which is exactly $(\UC + 1) n^{d-2}$. \qed 
\end{proof}

\begin{proof}[of  \Cref{lemma:l.integralitygap1}]
We use the above claim to upper bound the weight of an arbitrary feasible solution $F$ of \BPPT\ on $T_d$ by $(\UC + 1) n^{d-1}(1+\frac{1}{n})$. To this end, we do a case distinction based on the size of $S \coloneqq F \cap \{e_1, \cdots, e_n\},$ where $e_i$ is the edge connecting the root of $T_d$ to the root of the $i$-th copy $T_{d-1}^{(i)}$ in the construction of $T_d$.

In particular, we show the desired bound on the weight of $F$ by induction over $d$; clearly the bound holds for $d = 1$. \\ \\
\noindent\emph{Case 1: $S = \emptyset$.} In this case, $F \cap T_{d-1}^{(i)}$ is a feasible solution for $T_{d-1}^{(i)}$ for each $i \in [n]$. Thus, by induction, we have
\[
    w(F) \leq n \cdot (\UC + 1) n^{d-2}(1 + \frac{1}{n}) = (\UC + 1) n^{d-1}(1 + \frac{1}{n}).
\]
\emph{Case 2: $S = \{e_i\}.$} Note that there cannot exist a $j \in [n]$ and a maximal chain $Q \in \mathcal{Q}_{d-1}^{(j)}$ such that $|F \cap Q| \geq 2$, as otherwise the constraint corresponding to a commodity containing the path $Q \cup \{e_i, e_j\}$ is violated. Thus, by Claim (1), we have
\[
    w(F) \leq w(e_i) + n (\UC + 1) n^{d-2} = (\UC + 1) n^{d-2} + (\UC + 1) n^{d-1} = (\UC + 1) n^{d-1}(1 + \frac{1}{n}).
\]
\emph{Case 3: $|S| \geq 2$.} Let $e_i, e_j \in S$ for $i \neq j$. Then, we must have $F \cap T_{d-1}^{(i)} = F \cap T_{d-1}^{(j)} = \emptyset$, as otherwise, there would be a maximal chain $Q$ including one edge in $T_{d-1}^{(i)}$ and $e_i$, and thus the commodity with path $Q \cup \{e_j\}$ would be violated, analogously for $j$. Combining this with the argument from case 2, we get
\begin{multline*}
    w(F) \leq w(S) + (n-|S|) (\UC + 1) n^{d-2} \\
    = |S|(\UC + 1) n^{d-2} + (n-|S|)(\UC + 1) n^{d-2} = (\UC + 1) n^{d-1}.
\end{multline*}
This finishes the proof of the upper bound for the optimal solution value of (IP). \qed

\end{proof}

\begin{proof}[of \Cref{lemma:l.integralitygap2}]
Recall that, for the constructed instance $T_d$, we consider the fractional vector $x$ with values
\begin{itemize}
    \item $x(e) = \frac{1}{2}$ for every pendant edge, and
    \item  $x(e) = 3 \cdot 2^{j-d-1}$ for every depth-$j$ edge of $T_d$ for $j < d$.
\end{itemize}
Thus, the edges directly above the pendant edges receive $\frac{3}{4}$, and the values halve towards the root. To see that $x$ is feasible for (LP), it suffices to show feasibility for all commodities of type $\U_d$, and all newly added commodities in $\mathcal{C}_d$, since feasibility of all $\C$-type commodities which were copied from instance $T_{d-1}$ follows by induction. 

Each commodity in $\U_d$ contains exactly two pendant edges, so $x$ is clearly feasible with respect to $\U_d$.
For the $\C$ commodities, fix some commodity with path $\{e_i, e_j\} \cup Q_{d-1}^{(i)}$. The commodity contains a pendant edge with LP-value $\frac{1}{2}$. Additionally, the commodity contains one depth-$j$ edge for each $j \in \{ 2, \ldots, d-1\}$ and two depth-$1$ edges. The non-pendant edges carry a total LP-value of
\[
    2\cdot 3 \c 2^{1-d - 1} + \sum_{j=2}^{d-1} 3 \c 2^{j-d - 1} = 3\left(2^{1-d} + \left(\frac{1}{2} - 2^{1 - d}\right)\right) = \frac{3}{2}
\]
by geometric sum. Hence, $x$ is feasible on all commodities of type $\C$ which have a budget of $2$ each.

Since the pendant edges have a total weight of $(\UC + 1)^2 n^{d-1}$ and every other depth-level has a total weight of $(\UC + 1) n^{d-1}$, we conclude
\begin{multline*}
    w^{\intercal}x = \frac{1}{2}(\UC + 1)^2 n^{d-1}+ \left(\sum_{j=1}^{d-1} 3 \c 2^{j-d - 1}\right) (\UC + 1) n^{d-1} = \\
    (\UC + 1) n^{d-1}\left(\frac{\UC+1}{2} + 3 (2^{-1}- 2^{-d})\right). \qedhere
\end{multline*}
\end{proof}

We conclude this subsection with a proof of \Cref{thm:two_plus_uc_half_integrality_gap}, thereby establishing that the integrality gap of the standard LP relaxation relative to $\ucong$ matches the guarantee of our first approximation algorithm.

\begin{proof}[of \Cref{thm:two_plus_uc_half_integrality_gap}]
 Combining the two previous lemmas yields
\[
    \sup_{\mathcal I} \frac{\OPT_{\textsc{LP}}}{\OPT_{\textsc{IP}}} \geq \frac{(\UC + 1) n^{d-1} (\frac{\UC + 1}{2} + 3 \cdot (2^{-1} - 2^{-d}))}{(\UC + 1) n^{d-1}(1 + \frac{1}{n})} \overset{n, d \rightarrow\infty}{\longrightarrow} \frac{\UC + 1}{2}+ \frac{3}{2} = 2 + \frac{\UC}{2}.
\]
as a lower bound for the integrality gap. Exactness follows from \Cref{thm:two_plus_uc_half_approx}. \qedhere
\end{proof}

\subsection{$(1+\conge/2)$-approximation}
\label{app:flr_parameter_congestion_improved}

We present a variant of the algorithm from \Cref{sec:2_plus_uc_half} whose guarantee only depends on the congestion $\conge$ rather than $\ucong$. Clearly, for $\conge=1$ solving $\BPPT$ is trivial, and thus we assume $\conge \geq 2$ for the remainder of this section.

\flrCongestionImproved*

This result improves on \Cref{thm:two_plus_uc_half_approx} whenever $\conge \leq \ucong + 1$. As a consequence, the integrality gap of (LP) for a fixed value of $\conge$ lies in the interval $[\frac{\conge+1}{2}, \frac{\conge+2}{2}]$. Here, the lower bound is attained on a star with $\conge + 1$ unit weight edges, where each pair of outer vertices is connected by a unit-budget commodity.

The algorithm presented in this section differs from the one in \Cref{sec:2_plus_uc_half} (\Cref{alg:2_plus_uc_half}) only in the way the auxiliary instance $\I'$ is constructed. The core new observation is that the auxiliary instance does in fact not need to be an instance of $\BPPT$, and it is e.g. allowed to include commodities consisting of two disconnected path segments. All that is needed is that (a) each commodity of $\I$ is cut into at most $b_i$ unit-budget pieces in $\I'$, which ensures feasibility, and (b) we can find edges with sufficiently small local neighborhoods in the auxiliary instance. The remainder of the algorithm, that is, the application of fractional local ratio and the construction of a feasible solution via reverse greedy is unchanged from \Cref{sec:2_plus_uc_half}. \\

\noindent\textbf{Construction of the auxiliary instance $\I'$.} Given an instance $\I$ of $\BPPT$, let $x^*$ be an optimal fractional solution of the LP-relaxation. Choose an arbitrary vertex $r$ as the root of $T$. We keep unit-budget commodities unchanged. For each bent commodity $(P_i, b_i)$ of budget at least two, split $P_i$ into its two sides $L_i$ and $R_i$ at its apex. W.l.o.g., assume that $x^*(L_i) \geq x^*(R_i)$. For each vertical commodity $(P_i, b_i)$ of budget at least two we write $L_i \coloneqq P_i$ and $R_i \coloneqq \emptyset$. 
We now further split up all $L_i$ and $R_i$ into unit-budget pieces. Initialize $\mu \coloneqq 0$ and walk along $L_i$ and $R_i$ starting at the apex of $P_i$ for each, and keep a current piece of $x^*$-mass $\mu$. The next edge $e$ in $L_i$ (or $R_i$, respectively) is added to the current piece if $\mu \leq 1$ or $\mu + x^*(e) \leq 3/2$. Otherwise, cut off the current piece and start a new piece beginning with edge $e$ and mass $\mu = x^*(e)$. In other words, we begin a new piece if the mass of the current piece exceeds $1$ and adding a new edge would raise the mass above $3/2$.

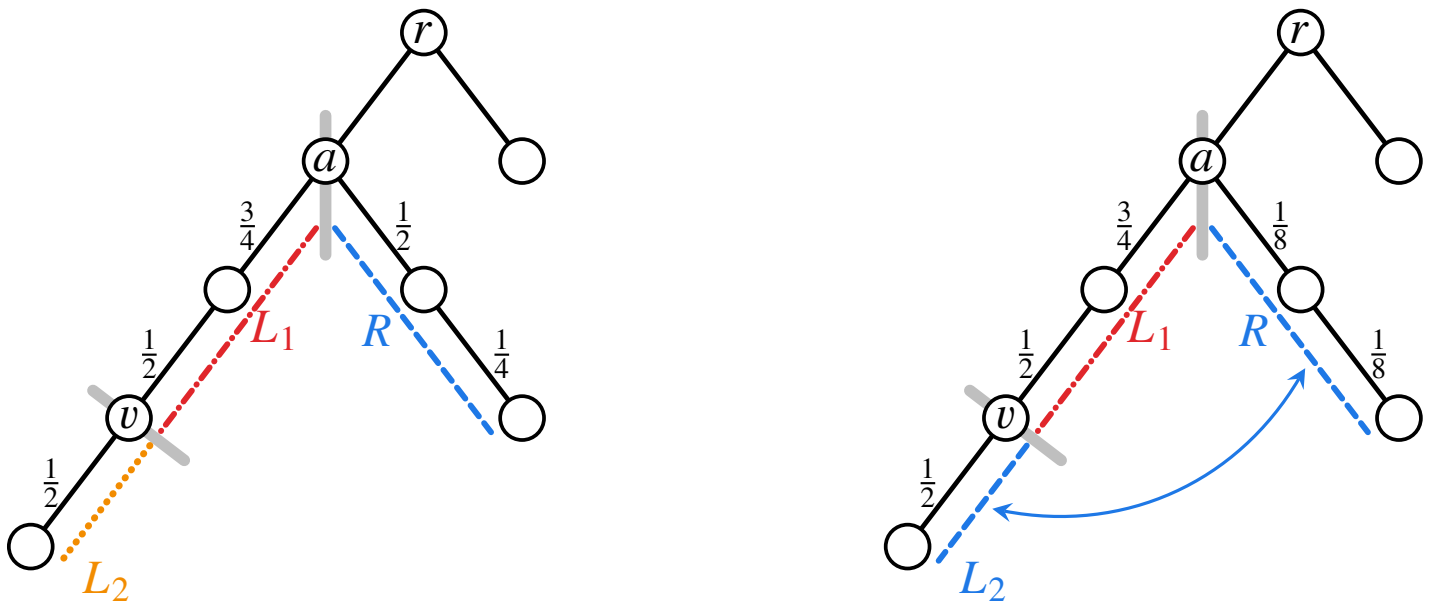
\begin{figure}[t]
\centering
\definecolor{bpptRed}{HTML}{E0262B}
\definecolor{bpptOrange}{HTML}{F28C00}
\definecolor{bpptBlue}{HTML}{1E78E6}
\tikzset{
  aux vertex/.style = {circle, draw, fill=white, line width=0.7pt,
                       minimum size=2.9mm, inner sep=0pt, font=\footnotesize},
  aux edge/.style   = {line width=0.9pt},
  aux xval/.style   = {font=\scriptsize, inner sep=2.4pt},
  aux cut/.style    = {line width=2.2pt, black!25, line cap=round},
  aux piece/.style  = {line width=1.1pt, line cap=round},
  aux dashdot/.style= {aux piece,
                       dash pattern=on 2.6pt off 1.4pt on 0.1pt off 1.4pt},
  aux dotted/.style = {aux piece, line width=1.3pt,
                       dash pattern=on 0.1pt off 1.8pt},
  aux dashed/.style = {aux piece, dash pattern=on 2.6pt off 1.6pt},
  aux merge/.style  = {line width=0.6pt, bpptBlue, <->, >=stealth},
  aux plabel/.style = {font=\footnotesize, inner sep=1pt},
}

\newcommand{\auxPiece}[6]{%
  \coordinate (aux s) at ($(#3)!#5!(#4)$);
  \coordinate (aux t) at ($(#4)!#6!(#3)$);
  \coordinate (aux o) at ($(aux s)!2.2mm!#2:(aux t)$);
  \draw[#1] (aux o) -- ($(aux t)+(aux o)-(aux s)$);}
\newcommand{\auxTree}[5]{%
  \coordinate (r)  at (0.65,0.85);
  \coordinate (a)  at (0,0);
  \coordinate (s)  at (1.3,0);
  \coordinate (u1) at (-0.65,-0.85);
  \coordinate (v)  at (-1.3,-1.7);
  \coordinate (u3) at (-1.95,-2.55);
  \coordinate (w1) at (0.65,-0.85);
  \coordinate (w2) at (1.3,-1.7);
  \draw[aux cut] (0,0.3) -- (0,-0.62);
  \draw[aux cut] ($(v)!-3mm!90:(u3)$) -- ($(v)!4.6mm!90:(u3)$);
  \draw[aux edge] (r) -- (a);
  \draw[aux edge] (r) -- (s);
  \draw[aux edge] (a)  -- node[aux xval, left]  {$#1$} (u1);
  \draw[aux edge] (u1) -- node[aux xval, left]  {$#2$} (v);
  \draw[aux edge] (v)  -- node[aux xval, left]  {$#3$} (u3);
  \draw[aux edge] (a)  -- node[aux xval, right] {$#4$} (w1);
  \draw[aux edge] (w1) -- node[aux xval, right] {$#5$} (w2);}
\newcommand{\auxVertices}{%
  \node[aux vertex] at (r) {$r$};
  \node[aux vertex] at (a) {$a$};
  \node[aux vertex] at (v) {$v$};
  \foreach \p in {s, u1, u3, w1, w2}{\node[aux vertex] at (\p) {};}}
\begin{tikzpicture}
  \begin{scope}
    \auxTree{\frac34}{\frac12}{\frac12}{\frac12}{\frac14}
    \auxPiece{aux dashdot, bpptRed}{90}{a}{v}{3.9mm}{0.6mm}
    \auxPiece{aux dotted, bpptOrange}{90}{v}{u3}{0.6mm}{0.5mm}
    \auxPiece{aux dashed, bpptBlue}{-90}{a}{w2}{3.9mm}{0.5mm}
    \auxVertices
    \node[aux plabel, bpptRed]    at (-0.34,-1.12) {$L_1$};
    \node[aux plabel, bpptOrange] at (-1.45,-2.78) {$L_2$};
    \node[aux plabel, bpptBlue]   at ( 0.34,-1.12) {$R$};
    \node[font=\small, anchor=base] at (-0.3,-3.35)
      {(a) $b = 3$, \ $x^*(R) = \frac34 \ge \frac12$};
  \end{scope}
  \begin{scope}[xshift=5.8cm]
    \auxTree{\frac34}{\frac12}{\frac12}{\frac18}{\frac18}
    \auxPiece{aux dashdot, bpptRed}{90}{a}{v}{3.9mm}{0.6mm}
    \auxPiece{aux dashed, bpptBlue}{90}{v}{u3}{0.6mm}{0.5mm}
    \auxPiece{aux dashed, bpptBlue}{-90}{a}{w2}{3.9mm}{0.5mm}
    \draw[aux merge] (-1.39,-2.30) to[bend right=35] (0.68,-1.37);
    \auxVertices
    \node[aux plabel, bpptRed]  at (-0.34,-1.12) {$L_1$};
    \node[aux plabel, bpptBlue] at (-1.45,-2.78) {$L_2$};
    \node[aux plabel, bpptBlue] at ( 0.34,-1.12) {$R$};
    \node[font=\small, anchor=base] at (-0.3,-3.35)
      {(b) $b = 2$, \ $x^*(R) = \frac14 < \frac12$};
  \end{scope}
\end{tikzpicture}
\caption{Construction of the auxiliary instance $\mathcal{I}'$ for a bent commodity $(P,b)$ with apex~$a$. The numbers on the edges are $x^*$-values, gray bars indicate cuts, and different colors indicate different commodities of~$\mathcal{I}'$.
In (a), a cut is placed at vertex $v$ since previously $\mu = \frac34 + \frac12 = \frac54 > 1$, and adding the next edge to the same piece would increase $\mu$ to $\frac74 > \frac32$. Note that the construction of the auxiliary instance from \Cref{sec:2_plus_uc_half_algo_descr} would not have placed the cut at $v$. We obtain exactly $3$ (and thus exactly $b$) pieces and thus, merging $L_2$ and $R$ is not necessary.
In contrast, in (b) the splitting procedure splits $P$ into $3 = b+1$ pieces. Thus, the last piece of $L$ and the single piece $R$ form a merged commodity, which is indicated by the blue arrow between the two pieces.}
\label{fig:construction_aux_instance}
\end{figure}

If this process splits up $P_i$ into at most $b_i$ pieces, each of the pieces becomes a unit-budget commodity in $\I'$. Otherwise, (as we will show) $L_i$ is split up into exactly $b_i$ pieces, and $R_i$ is a single piece. In this case we take the union  of $R_i$ and the last piece of $L_i$ and add this \emph{merged commodity} to $\I'$. We leave the remaining pieces of $L_i$ unchanged and add them to $\I'$ as unit-budget commodities to obtain exactly $b_i$ commodities in total. We continue this process for each commodity $i \in [m]$, and let $y \coloneqq x^*/2$. We make several observations regarding the structure of the constructed instance $\I'$. The splitting process used to construct the auxiliary instance is visualized in \Cref{fig:construction_aux_instance}.

\begin{lemma}
\label{lem:conge_at_most_bi}
    The process described above partitions each commodity $P_i$ into at most $b_i$ commodities in $\I'$. Hence, every feasible solution for $\I'$ is also feasible for $\I$, and every edge lies in at most $\conge$ commodities of $\I'$.
\end{lemma}
\begin{proof}
    Fix a commodity $P_i$, and let $Z\in\{L_i, R_i\}$ be a side of $P_i$ that gets split up into $q_Z \geq 2$ pieces $Z_1, \ldots, Z_{q_Z}$ with $Z_1$ being the topmost piece and $Z_{q_Z}$ being the bottommost piece. Assume that an edge $e$ is the first edge of piece $Z_{j+1}$. By construction, we must have $x^*(Z_j) > 1$ and $x^*(Z_j) + x^*(e) > \frac32$. This implies $x^*(Z_j) > 1$ for all $j < q_Z$ and $x^*(Z_{q_Z-1}) + x^*(Z_{q_Z}) > \frac{3}{2}$, since otherwise piece $Z_{q_Z}$ would not have been started. Summing up, $x^*(Z) > q_Z - \frac{1}{2}$. Note that if $q_{L_i} = q_{R_i} = 1$, then we have $q_{L_i} + q_{R_i} = 2 \leq b_i$, and thus w.l.o.g. we consider the case where at least one of $q_{L_i}$ and $q_{R_i}$ is at least two. We distinguish cases based on the LP weight of $R_i$.

    \emph{Case 1: $x^*(R_i) \geq \frac12$.} Recall that we assumed $x^*(L_i) \geq x^*(R_i)$. Thus, by the observation above it holds that $q_Z \leq x^*(Z) + \frac12$ for both sides $Z \in \{L_i, R_i\}$ and this inequality is strict when $q_Z \geq 2$. As we have $q_Z \geq 2$ for at least one of the two sides $L_i$ and $R_i$, we obtain $q_{L_i} + q_{R_i} < x^*(P_i) + 1 \leq b_i + 1$, which implies $q_{L_i} + q_{R_i} \leq b_i$ by integrality.

    \emph{Case 2: $x^*(R_i) < \frac12$.} In this case, $q_{R_i} \leq 1$ by construction of the auxiliary instance. Thus, we must have $q_{L_i} \geq 2$ which implies $q_{L_i} < x^*(L_i) + \frac12 \leq b_i + \frac{1}{2}$, so $q_{L_i} \leq b_i$ by integrality. Summing up, there are at most $b_i + 1$ pieces, and in case there are exactly $b_i + 1$, two of them are merged during the construction, and thus $P_i$ is split up into exactly $b_i$ commodities in $\I'$.

    In both cases, $P_i$ is partitioned into at most $b_i$ commodities in $\I'$. A feasible solution for $\I'$ contains at most one edge from each of them, and thus at most $b_i$ edges from $P_i$ in total. This shows feasibility for $\I$. Also note that the congestion $\conge$ remains unchanged when considering $\I'$ rather than $\I$, since $\I'$ partitions the commodities of $\I$. \qedhere
\end{proof}
\newpage
\begin{observation}
\label{obs:conge_structure_commodities}
Each commodity $P$ of the auxiliary instance $\I'$ is of one of the following three types.
\begin{itemize}
    \item[(U)] $P$ is a path with $y(P) \leq \frac12$,
    \item[(V)] $P$ is a \emph{vertical} path with bottommost edge $e$, and we have $y(P) \leq \frac34$ or $y(P \setminus \{e\}) \leq \frac12$,
    \item[(M)] $P = L \cup R$ is a \emph{merged commodity} consisting of two edge-disjoint (possibly empty) paths and we have $y(P) \leq \frac12$ and $y(R) \leq \frac14$. Furthermore, the highest vertex of $R$ is an ancestor of the highest vertex of $L$.
\end{itemize}
\end{observation}
\begin{proof}
    Commodities of type (U) arise from unit-budget commodities in $\I$ for which we have $y(P) \leq 1/2$ since $x^*$ is feasible for $\I$. Commodities of type (V) are vertical pieces constructed during the cutting process that were not merged together with another piece. 
    Assume towards a contradiction that for one such piece neither of the two inequalities $y(P) \leq \frac{3}{4}$ and $y(P \setminus \{e\}) \leq \frac{1}{2}$ holds. Thus, using $y = \frac{x^*}{2}$, when constructing the piece, we had $\mu > 1$ before adding the last edge $e$ since $y(P \setminus \{e\}) > \frac12$, and $\mu + x^*(e) > \frac32$ since $y(P) > \frac34$. This contradicts the way the auxiliary instance is constructed, since we would have created a new piece beginning at $e$. 
    Commodities of type (M) are the merged commodities which only occur from case 2 in \Cref{lem:conge_at_most_bi}, i.e., when $x^*(R) < \frac12$ and thus $y(R) \leq \frac14$. Note that in case a merged commodity $P$ is formed from an original commodity $P_i$, all $(b_i-1)$ commodities $(Z_1, \ldots, Z_{b_{i-1}})$ other than $P$ have $x^*$-mass at least 1 by construction of the auxiliary instance. Thus, $x^*(P) = x^*(P_i) - \sum_{j=1}^{b_i-1} x^*(Z_j) \leq b_i - (b_i-1) = 1$ which gives $y(P) \leq \frac12$. \qedhere
\end{proof}

We write $\mathcal{U}, \mathcal{V}$ and $\mathcal{M}$ to refer to the sets of commodities of type (U), (V) and (M) in $\I'$, respectively. During the run of the algorithm we apply fractional local ratio first to instance $\I'_0 \coloneqq \I'$ but later also to contracted instances $\I'_t$ which arise during the run of the remainder of \Cref{alg:2_plus_uc_half} (after the construction of the auxiliary instance). Observe that the properties from \Cref{lem:conge_at_most_bi} and \Cref{obs:conge_structure_commodities} hold also for these intermediate instances $\I'_t$, which allows us to use the proof of \Cref{thm:two_plus_uc_half_approx} (cf. \Cref{sec:2_plus_uc_half_algo_analysis}) verbatim once we have shown the existence of an edge with a sufficiently small neighborhood in every $\I'_t$. \\

\noindent\textbf{Small neighborhoods.} It remains to prove a lemma replacing \Cref{lem:cheap_neighborhood_edge} w.r.t. the new construction of the auxiliary instance, and parameterized in $\conge$ instead of $\ucong$. Recall that we define the neighborhood $N_{\I}[e]$ of edge $e$ in instance $\I$ to be $\{e\}$ together with the set of edges $f$ that share a commodity with $e$ in $\I$.

\begin{lemma}
\label{lem:small_local_neighborhood_conge_improved}
Let $T = (V, E)$ be a tree with root $r$ and let $y \in [0, \frac12]^{E}$ and $\conge \geq 2$. Furthermore, let $\I$ be a family of nonempty subsets of $E$, each of which is of one of the three types (U), (V) or (M) from \Cref{obs:conge_structure_commodities} with respect to $y$, and such that each edge is part of at most $\conge$ subsets in $\I$. Then, there exists an edge $e\in E$ with
\[y(N_{\I}[e]) \leq \frac12 + \frac{\conge}{4}.\]
\end{lemma}

\begin{proof}
    Recall the following notions which are unchanged from the proof of \Cref{lem:cheap_neighborhood_edge}. We write $N[e]$ rather than $N_\I[e]$. For vertices $u, v \in V$ we denote by $P_{u,v}$ the unique $u$--$v$--path in $T$. If $u$ is a descendant of $v$, we say that $u$ is \emph{lower} than $v$ (and that $v$ is \emph{higher} than $u$). A vertex $v$ is called \emph{heavy} if there is a vertex $u$ that is lower than $v$ such that $y(P_{u,v})\ge \tfrac{1}{4}$. 
    We select a leaf $\ell \in V$ as follows, and let $e \in E$ denote the unique edge incident to $\ell$.
    \begin{itemize}
        \item If no vertex is heavy, $\ell$ is any leaf.
        \item Otherwise, let $z$ be a lowest heavy vertex, and let $\ell$ be a leaf such that $y(P_{\ell,z})\geq \tfrac{1}{4}$. 
    \end{itemize}
    We will need the following property of this choice of $e$, which we will refer to as $(\star)$: For every heavy vertex $v$ on the path from $r$ to $\ell$ we have $y(P_{v,\ell}) \geq \frac14$. \medskip

   Let $\I_e \subseteq \I$ be the set of subsets that contain edge $e$. Clearly, we have $|\I_e| \leq \conge$. The case $\I_e = \emptyset$ is trivial, so assume that the set is nonempty for the remainder of the proof. Each commodity $P \in \I_e$ can be split up into a (maximal) vertical part $P_{t_P, \ell}$ that includes edge $e$ and a \emph{far part} $B_P$ that includes the rest of $P$. We call $t_P$ the \emph{top} of commodity $P$.
   \begin{itemize}
       \item If $P \in \mathcal{U}$, then $B_P$ is the vertical path from $t_P$ to $v_P$ where $v_P$ is the second endpoint (other than $\ell$) of $P$.
       \item If $P \in \mathcal{V}$, then $B_P = \emptyset$.
       \item If $P \in \mathcal{M}$, then $B_P$ is the other one of the two disjoint components of $P$ that $e$ is not a part of.
   \end{itemize}
   In essence, the top $t_P$ of a commodity fulfills a similar role to the apex $a_P$ in the proof of \Cref{lem:cheap_neighborhood_edge}, and the far part $B_P$ serves the purpose of the apex-to-endpoint path $P_{a_P, v_P}$ in said proof.
   
   Let $P^*$ be the commodity whose top $t^* = t_{P^*}$ is highest among all commodities $P \in \I_e$. We have $P_{t_P, \ell} \subseteq P_{t^*, \ell}$ for all commodities $P \in \I_e$, and thus we can split the $y$-mass on $N[e]$ as follows:
   \[y(N[e]) \leq y(P_{t^*, \ell}) + \sum_{P \in \I_e} y(B_P).\]
   We bound the $y$-mass on the far parts $B_P$. To this end, we prove the following claim:

    \begin{claim}[2]\label{cl:claimOnePointFive}
    For every $P \in \I_e$ we have $y(B_P) \leq \frac12 - s$ where $s \coloneqq \max\{y(e), \frac{1}{4}\}$.
    \end{claim}

    \begin{proof}
    If $P \in \mathcal{V}$ we have $B_P =\emptyset$, so we only need to consider the cases $P \in \mathcal{U}$ and $P \in \mathcal{M}$; in either case we have $y(P) \leq \frac12$. We always have $y(P_{t_P, \ell}) \geq y(e)$ and thus $y(B_P) \leq \frac12 - y(e)$. It remains to show that we also have $y(B_P) \leq \frac14$.
    If $t_P$ is heavy, this is clear since $y(P_{t_P, \ell}) \geq \frac14$ by $(\star)$, which implies $y(B_P) \leq \frac12 - y(P_{t_P, \ell}) \leq \frac14$. Thus, assume that $t_P$ is not heavy.
    \begin{itemize}
    \item If $P$ is of type (U), then $B_P$ is contained in a vertical path down from $t_P$ which gives $y(B_P) \leq \frac14$.
    \item If $P = L \cup R$ is of type (M) and $e \in R$, then the highest vertex from $L$ is lower than $t_P$, and thus $y(B_P) = y(L) \leq \frac14$ since $t_P$ is not heavy.
    \item Lastly, if $P = L \cup R$ is of type (M) and $e \in L$ we have $y(B_P) = y(R) \leq \frac14$ by \Cref{obs:conge_structure_commodities}.
    \end{itemize}
    This finishes the proof of the claim. \qedhere
    \end{proof}

    If $P^*$ is of type (U) or (M), we have $y(P_{t^*, \ell}) + y(B_{P^*}) = y(P^*) \leq \frac12$. This gives
    \[y(N[e]) \leq y(P_{t^*, \ell}) + \sum_{P \in \I_e} y(B_P) \leq \frac12 + \sum_{P^* \neq P \in \I_e} y(B_P) \leq \frac12 + (\conge - 1)\frac14 \leq \frac12 + \frac{C}{4}.\]
    If $P^*$ is of type (V) we have $y(P^*) \leq \frac12 + s$ as otherwise the algorithm would have placed a cut before edge $e$.
    Combined with Claim (2), we obtain
    \[
    y(N[e]) \leq 
    y(P_{t^*, \ell}) + \sum_{P \in \I_e} y(B_P) \leq \left(\frac12 + s \right) + (\conge-1)\left(\frac12 - s\right) \leq 
    \frac12 + \frac{C}{4}.\qedhere
    \]
\end{proof}

\noindent\textbf{Applying Fractional Local Ratio.} Using \Cref{lem:small_local_neighborhood_conge_improved}, we can now prove \Cref{thm:flrCongestionImproved}. Compared to the $2 + \ucong/2$-approximation given in \Cref{alg:2_plus_uc_half}, only the first phase of the algorithm (building the auxiliary instances) in lines 1 -- 12 changes, the application of fractional local ratio and the construction of $F_{\ALG}$ via reverse greedy stay unchanged.

\begin{proof}[of \Cref{thm:flrCongestionImproved}]
    Clearly, the construction of the auxiliary instance and application of fractional local ratio take only polynomial time. The auxiliary instance $\I'_0 \coloneqq \I'$ as well as all later instances $\I'_t$ constructed by contracting edges and commodities in $\I'_{t-1}$ fulfill the structural properties of \Cref{obs:conge_structure_commodities} and thus, \Cref{lem:small_local_neighborhood_conge_improved} is applicable in every iteration $t$. As the algorithm chooses edge $e_t$ which minimizes $y(N_{\I'_t}[e_t])$ among all edges in $\I'_t$ this gives
    \[y(N_{\I'_t}[e_t]) \leq \frac12 + \frac{\conge}{4}\]
    for all $t$. The proof of \Cref{thm:two_plus_uc_half_approx} (cf. Appendix \ref{appendix:proof_two_uc_half_apx}) now applies verbatim when exchanging the bound $1 + \ucong/4$ with $1/2 + \conge/4$ throughout. This yields the desired approximation guarantee. \qedhere
\end{proof}

\section{$\tfrac{\mincap}{\mincap -1}$-approximation on Instances with  $\mincap \geq 2$}

\subsection{Pseudocode}\label{appendix:ell_ell_minus_one_pseudocode} 

Pseudocode of the algorithm attaining an approximation guarantee of $\tfrac{\mincap}{\mincap-1}$ for $\mincap \geq 2$ is shown in \Cref{alg:LP_guided_splitting}.

\begin{algorithm}
    \caption{LP-guided splitting}
    \label{alg:LP_guided_splitting}
    
    \textbf{Input:} An instance $\I$ consisting of a tree $T$, commodities $(P_i, b_i)_{i \in [k]}$ and weights $(w_e)_{e\in E}$ with $b_i \geq 2$ for all $i \in [m]$. Assume $\I$ contains at least one bent commodity.\\
    \textbf{Output:} A feasible solution $F_{\ALG}.$
    \begin{algorithmic}[1]
    
        \State Compute an optimal fractional solution $x^*$ to (LP).
        \State Choose an arbitrary root $r \in V$.
        \State $\alpha := \max_{i \in [m]}\{\frac{b_i}{b_i - 1} \mid P_i \text{ bent}\}$
        \For{$i\in[m]$ bent}
            \State Split $(P_i, b_i)$ into two commodities with paths $L_i$ and $R_i$.
            \State Set $b_{L_i} = \lceil x^*(L_i)/\alpha \rceil$ and $b_{R_i} = \lceil x^*(R_i)/\alpha \rceil$.
        \EndFor
        \State Compute an optimal integral solution $F_{\ALG}$ for the resulting instance.
        \State \Return $F_{\ALG}$
        
    \end{algorithmic}
\end{algorithm}

\subsection{Integrality Gap}\label{appendix:ell_ell_minus_one_integrality_gap}

The following proof of \Cref{thm:ell_ell_minus_one_integrality_gap} is very similar to the proof of \Cref{thm:two_plus_uc_half_integrality_gap}. We nonetheless present it in full in order to have a self-contained proof of the statement.

\EllEllMinusOneIntegralityGap*

\begin{proof}
We provide a construction that proves the lower bound. For this, we fix some $n \geq 2$ and recursively define a family of rooted trees $T_d$ for $d \in \{1,2, \ldots\}$ as follows:
\begin{itemize}
    \item $T_1$ is a path of length $\mincap - 1$ rooted at either one of its endpoints.
    \item For $d \geq 2$, we construct $T_d$ by taking a root $r$ and $n$ copies $T_{d-1}^{(1)},\ldots,T_{d-1}^{(n)}$. For $i\in[n]$, we connect $r$ and the root of the $i$-th copy $T_{d-1}^{(i)}$ with an edge $e_i$.
\end{itemize}
We say that a \emph{maximal chain} of $T_d$ is a root-to-leaf path of $T_d$ and denote the set of all maximal chains of $T_d$ by $\mathcal{Q}_d$. Similarly, $\mathcal{Q}_{d-1}^{(i)}$ denotes the set of maximal chains in the $i$-th copy of $T_{d-1}$ used in the construction of $T_d$. We refer to the edges that are part of one copy of $T_1$ as \emph{path edges}, that is, the the path edges are exactly the edges whose depth is at least $d$. We refer to all other edges as \emph{upper edges}.
We define the edge weights $w_e$ on $T_d$ to be the number of leaves below edge $e$ (when directing all edges away from the root $r$). That is, for an edge $e$ at the $j$-th depth level we have
\[
    w_e \coloneqq |\{Q \in \mathcal{Q}_d \ | \ e \in Q\}| = 
    \begin{cases}
        1, & e \text{ is a path edge}, \\
        n^{d-1-j}, & e \text{ is an upper edge}.
    \end{cases}
\]
Using these weights, the total weight of each level of $T_d$ is exactly the number of leaves in $T_d$, that is, $n^{d-1}$.
Next, we define the set of commodities $\C_d$. We set $\C_1 = \emptyset$. For $d > 1$, the set $\C_d$ contains the following commodities, with $b \equiv \mincap$:
\begin{itemize}
    \item The union of all commodities $\C_{d-1}^{(i)}$ and
    \item the new commodities $\{\{e_i, e_j\} \cup Q \ | \ i \neq j, \ Q \in \mathcal{Q}_{d-1}^{(i)}\}$ consisting of one edge adjacent to $r$ and a maximal chain of $T_d$.
\end{itemize}
The proof is now an immediate consequence of the following two lemmas.

\begin{lemma}
\label{lemma:integrality_gap_beta_2}
The weight of any integral solution for \BPPT\ on $T_d$ can be upper bounded by $(\mincap - 1) n^{d-1} + n^{d-2}$.
\end{lemma}
\begin{lemma}
\label{lemma:integrality_gap_beta_1}
Given instance $T_d$ as constructed above, set $x(e) = 1$ for every path edge, and $x(e) = 2^{j-d}$ for every depth-$j$ edge of $T_d$ for $j < d$.
Then $x$ is a feasible LP-solution of
 weight at least $n^{d-1}(\mincap-2^{1-d})$.
\end{lemma}
Combining the two bounds yields the integrality gap
\[\sup_{\mathcal I} \frac{\OPT_{\textsc{LP}}}{\OPT_{\textsc{IP}}} \geq \frac{n^{d-1}(\mincap - 2^{1-d})}{(\mincap-1)n^{d-1} + n^{d-2}} \overset{n, d \rightarrow\infty}{\longrightarrow} \frac{\mincap}{\mincap-1}\]
as a lower bound, and the upper bound follows from \Cref{thm:ell_ell_minus_one_approx}. \qed
\end{proof}
For the proof of \Cref{lemma:integrality_gap_beta_2}, the following claim turns out to be helpful.

\begin{claim}[3]\label{claim2}
    Let $F'$ be any integral solution of $\BPPT$ on $T_d$ such that $F' \cap T_{d-1} ^{(i)}$ intersects each maximal chain $Q \in \mathcal{Q}_{d-1}^{(i)}$ of the subtree $T_{d-1}^{(i)}$ at most $\mincap - 1$ times. Then, we have $w(F' \cap T_{d-1} ^{(i)}) \leq (\mincap - 1)n^{d-2}$.
\end{claim}

\begin{proof}[of Claim (3)]
We assume $d >2$ since the case $d \leq 2$ is trivial. Given a solution $F'$ satisfying the property above, we push all its edges in $T_{d-1}^{(i)}$ down into the $\mincap-1$ lowest levels: We replace each edge $e \in F' \cap T_{d-1}^{(i)}$ above these levels by one path edge $e \not\in F'$ on each maximal chain $Q\in\mathcal{Q}_{d-1}^{(i)}$ containing $e$.
In this process, (i) the number of cuts on each maximal chain stays constant, (ii) the weight of the solution stays constant, since $w(e)$ is the number of leaves under $e$, and each of the corresponding path edges has weight $1$, and (iii) the solution remains feasible. Recall that the cost of every level in $T_{d-1}^{(i)}$ is exactly $n^{d-2}$. Thus, the weight of $F' \cap T_{d-1}^{(i)}$ is at most the weight of the $\mincap-1$ lowest levels of $T_{d-1}^{(i)}$ which is exactly $(\mincap-1)n^{d-2}$. \qed
\end{proof}

\begin{proof}[of \Cref{lemma:integrality_gap_beta_2}] 
We use the above claim to upper bound the cost of an arbitrary feasible solution $F$. To this end, we do a case distinction based on the size of $S \coloneqq F \cap \{e_1, \cdots, e_n\}$, where $e_i$ is the edge connecting the root of $T_d$ to the root of the $i$-th copy $T_{d-1}^{(i)}$ in the construction of $T_d$.

In particular, we show the desired bound on the weight of $F$ by induction over $d$; clearly the bound holds for $d = 1$. Now, let $d > 1$.\\

\noindent\textit{Case 1: $S = \emptyset$.} In this case, $F \cap T_{d-1}^{(i)}$ is a feasible solution for $T_{d-1}^{(i)}$ for each $i \in [n]$. Thus, by induction, we have
\[w(F) \leq n \cdot ((\mincap - 1)n^{d-2} + n^{d-3}) = (\mincap-1)n^{d-1}+n^{d-2}.\]

\noindent\textit{Case 2: $S = \{e_i\}.$} Note that there cannot exist a $j \in [n]$ and a maximal chain $Q \in \mathcal{Q}_{d-1}^{(j)}$ such that $|F \cap Q| \geq \mincap$, as otherwise the constraint corresponding to a commodity with path $Q \cup \{ei, e_j\}$ is violated. Thus, by Claim (3), we have
\[
    w(F) \leq w(e_i) + n \cdot (\mincap-1)n^{d-2} = (\mincap-1)n^{d-1} + n^{d-2}.
\]

\noindent\textit{Case 3: $|S|\geq 2$.}
Since $w_e=|\{Q\in\mathcal{Q}_d \mid e\in Q\}|$ for every edge $e$, double counting gives
\[
  w(F)=\sum_{e\in F} w_e=\sum_{Q\in\mathcal{Q}_d}|F\cap Q| .
\]
Every maximal chain of $T_d$ is of the form $Q= \{ e_k \} \cup Q'$ with $i \in [n]$ and
$Q' \in \mathcal{Q}^{(i)}_{d-1}$. Since $|S|\geq 2$, there is some $e_j\in S$ with $j \neq k$.
The commodity with path $\{e_k,e_j\} \cup Q'$ has budget $\mincap$ and contains $e_j \in F$,
hence $|F \cap Q| \leq \mincap-1$. Summing over all $|\mathcal{Q}_d|=n^{d-1}$ maximal chains yields $w(F) \leq (\mincap-1)n^{d-1}$.
  
This finishes the proof of the upper bound for the optimal solution value of (IP). \qed

\end{proof}

\begin{proof}[of \Cref{lemma:integrality_gap_beta_1}] 
Recall that, for the constructed instance $T_d$, we consider the fractional vector $x$ with values
\begin{itemize}
    \item $x(e) = 1$ for every path edge, and
    \item  $x(e) = 2^{j-d}$ for every depth-$j$ edge of $T_d$ for $j < d$, that is, every upper edge.
\end{itemize}

To see that $x$ is feasible suffices to show feasibility for all newly added commodities of $\mathcal{C}_d$, since feasibility for all copied commodities follows by induction.

Fix some commodity with path $\{e_i, e_j\} \cup Q$. The commodity contains $\mincap-1$ path edges of total LP-value $\mincap-1$. Additionally, the commodity contains one depth-$j$ upper edge for each $j = 2, \ldots, d-1$ and two depth $1$ upper edges. The upper edges carry a total LP-value of
\[2\cdot 2^{1-d} + \sum_{j=2}^{d-1} 2^{j-d} = 2^{2-d} + (1-2^{2-d}) = 1\]
by geometric sum. Hence, the solution is feasible.
Since each depth-level has a total weight of $n^{d-1}$, we have
\[\OPT_{\LP}(T_d) \geq w^{\intercal}x = n^{d-1}\left((\mincap - 1) + \sum_{j=1}^{d-1} 2^{j-d}\right) = n^{d-1}(\mincap - 2^{1-d}).\qedhere \]
\end{proof}

\section{A $\lceil (\conge+1)/3 \rceil$-approximation in $\XP$-time}
\label{sec:coloring_approx}

Recall that $\BPPT$ instances where $T$ is a star and $b=\one$ correspond to instances of the weighted independent set problem (MWIS) in a graph of maximum degree $\Delta = \conge$. For MWIS in graphs of maximum degree $\Delta$, Halldórsson and Lau \cite{HalldorssonLau97} use a coloring-based approach to obtain a simple $\lceil (\Delta + 1)/3 \rceil$-approximation, and later improve this guarantee to a more involved $(\Delta + 2)/3$-approximation. In this section, we extend the $\lceil (\Delta + 1)/3 \rceil$-approximation to $\BPPT$. While their approach for MWIS is very straightforward, adapting it to $\BPPT$ requires considerably more effort and a higher running time. For a given instance of $\BPPT$, let $\OPT$ be the weight of an optimal solution, and $b_{\max}:=\max\{b_i\mid i\in [m]\}.$

\begin{restatable}{theorem}{ColoringApproach}
\label{thm:congestion_approx_halldorsson_lau}
Given an instance of $\BPPT$ with tree $T=(V,E)$, we can compute a feasible solution $F$ with $w(F) \geq \OPT / \lceil (\conge + 1)/3 \rceil$ in time $(b_{\max} + 1)^{\mathcal O(C^2)} \cdot n^{\mathcal O(1)}$. Here, $n$ denotes the number of edges of tree $T$.
\end{restatable}

We can assume $b_{\max}<n$, and thus the running time is in $n^{O(C^2)}$. In particular, for $b=\one$ it is $2^{O(C^2)}\cdot n^{O(1)}$ and for $C\leq 2$ the algorithm is exact. For $C\geq 4$ we have $\lceil(C+1)/3\rceil<(C+1)/2$, so \Cref{thm:congestion_approx_halldorsson_lau} beats local search (cf. \Cref{thm:kexchange}) and every guarantee relative to~(LP), whose integrality gap is at least $(C+1)/2$ already on stars. \\

\noindent\textbf{General Idea.} We compute an edge coloring $c: E \rightarrow [q]$ with $q \coloneqq \lceil (\conge +1)/3\rceil$, and denote by $E = E_1 \cup \ldots \cup E_q$ the partition into color classes induced by $c$. We say that $\mathcal{I}_j$ is the instance obtained from $\mathcal{I}$ by contracting all edges $e$ with $c(e) \neq j$. That is, the feasible solutions to $\mathcal{I}_j$ are exactly those feasible solutions of $\mathcal{I}$ that are contained in $E_j$. We compute an optimal solution $F_j$ to each of the instances $\I_j$ and return the best of these candidate solutions. Let $F^*$ be an optimal solution to the original instance $\I$. Since $F^* \cap E_j$ is feasible for $\I_j$ for all $j \in [q]$, we have
\begin{equation}
\label{eq:coloring_approximation}
    w(F^*) = \sum_{j=1}^q w(F^* \cap E_j) \leq \sum_{j=1}^q w(F_j) \leq q \cdot \max_{j \in [q]} w(F_j)
\end{equation}
and thus, the algorithm computes a $q$-approximation. \\

Similar to Halldórsson and Lau \cite{HalldorssonLau97}, we make use of the following lemma, which goes back to Lovász.

\begin{lemma}[Lovász \cite{Lovsz1966DecompositionLemma}]\label{l.Lovasz}
    Let $G=(V,E)$ be an undirected graph of maximum degree $\Delta$. For $q$ with $3q \geq \Delta + 1$, there exists a partition $V = V_1 \cup \ldots \cup V_q$ into $q$ disjoint sets (``color classes'') such that each induced subgraph $G[V_j]$ has maximum degree at most 2. Moreover, such a partition can be found in polynomial time.
\end{lemma}

\noindent\textbf{The Approach of Halldórsson and Lau \cite{HalldorssonLau97}.} We briefly sketch the approach used to approximate an instance of maximum weighted independent set consisting of a graph $G=(V,E)$ and weights $w: V \rightarrow \R_+$. Using the Lemma above, the vertex set is partitioned into $q\ge \frac{1}{3}(\Delta +1)$ color classes $V = V_1 \cup \ldots \cup V_q$ 
 such that each induced subgraph $G[V_j]$ has maximum degree at most 2, so decomposes into disjoint paths and cycles. Thus, a max weight stable set $I_j$ for each of the $q$ subgraphs $G[V_j]$ can be found efficiently. The best candidate solution among all $I_j$ is clearly an independent set in $G$, and is a $q$-approximation. \\

\noindent\textbf{Adapting the approach for $\BPPT$.} For $\BPPT$ with $b=\one$, the analogous graph to the approach of Halldórsson and Lau would be the \emph{conflict graph} on~$E$, in which two edges are adjacent if, and only if, they lie on a common commodity. Its degree is not bounded in terms of~$C$, since every commodity forms a clique.

Instead, we use the fact that a bottom-up dynamic program over~$T$ only faces a real choice at the apex of a bent commodity, where it decides how the budget is shared between the two subtrees that the commodity enters. So we partition, at every vertex~$v$, the children of~$v$ with respect to the \emph{link graph}~$H_v$ (defined below), which has maximum degree at most~$C$. On a star rooted at its center, $H_v$ is the MWIS graph, and our algorithm coincides with that of Halldórsson and Lau. However, on general trees, the color of some edge $\{v,z\}$ (with $v$ being higher than $z$) influences not only decisions made at $v$, but also at every higher-up vertex $v'$ that is the apex of some bent commodity containing $\{v,z\}$. We resolve this by letting such \emph{pinned} edges inherit the color of their parent edge. The price is that a class is no longer a union of paths and cycles: at every vertex, it may contain up to~$C$ further vertices, the pinned children. We show that we still can solve the constructed classes exactly in time $(b_{\max}+1)^{O(C^2)}\cdot n^{\mathcal{O}(1)}$. \\

\subsection{Construction of the Edge Coloring}
\label{sec:constr_edge_coloring}
Given an instance of $\BPPT$ with the underlying tree $T=(V,E)$, fix a root $r$ of $T$, for $v \neq r$ denote by $e_v$ the edge between $v$ and its parent, and set $q \coloneqq \lceil (\conge + 1)/3 \rceil$. Recall that every bent commodity with path $P_i$ is the union of two vertical paths starting at the apex of the commodity, which we will call the two \emph{sides} of $P_i$, and say that $P_i$ is a \emph{link} at its apex. We introduce the \emph{link graph $H_v$ at $v$}. The vertices of this graph are the children of $v$, and two vertices $u_1$ and $u_2$ are adjacent in $H_v$ if there exists a link at $v$ containing both corresponding edges $\{v, u_1\}$ and $\{v, u_2\}$. Clearly, $H_v$ has a maximum degree of at most $\conge$. We say that an edge in $T$ is \emph{pinned} if it is on either side of some bent commodity, but not one of the two top edges of this commodity. By $Z_v$ we denote the set of children of $v$ for which $\{v,z\}$ is pinned. Hence, we have $Z_r = \emptyset$.

To construct the coloring $c: E \rightarrow [q]$, process the non-leaf vertices of $T$ from top to bottom. At vertex $v$, apply Lemma \ref{l.Lovasz} with $q\geq \frac{1}{3}(\conge +1)$ to the graph $H_v - Z_v$. As the maximum degree of that graph is at most $\conge$  this yields a partition of the children of $v$ that are not in $Z_v$ into at most $q$ sets, inducing a partitioning (coloring) of the edges in the set $\tilde{E}_v:=\{\{v,u\}\mid u \text{ child of } v \text{ with } u\not\in Z_v\}$ into $q$ color classes. For each $j \in [q]$, assign color $j$ to the edges in the $j$-th of these sets for every $v\in V$. Finally, all pinned edges in $Z_v$ inherit the color of their parent edge. That is, for every $z \in Z_v$, color the pinned edge $\{v,z\}$ with the color of $v$'s (unique) parent edge $e_v$. Let $c:E\to [q]$ be the resulting coloring. Clearly, this coloring $c$ is well-defined and can be computed efficiently; see \Cref{fig:construction_colouring} for an example. To solve the monochromatic instances $\I_j$ exactly, we make use of the following properties of the coloring.

\definecolor{colOne}{RGB}{0,0,179}
\definecolor{colTwo}{RGB}{230,115,0}
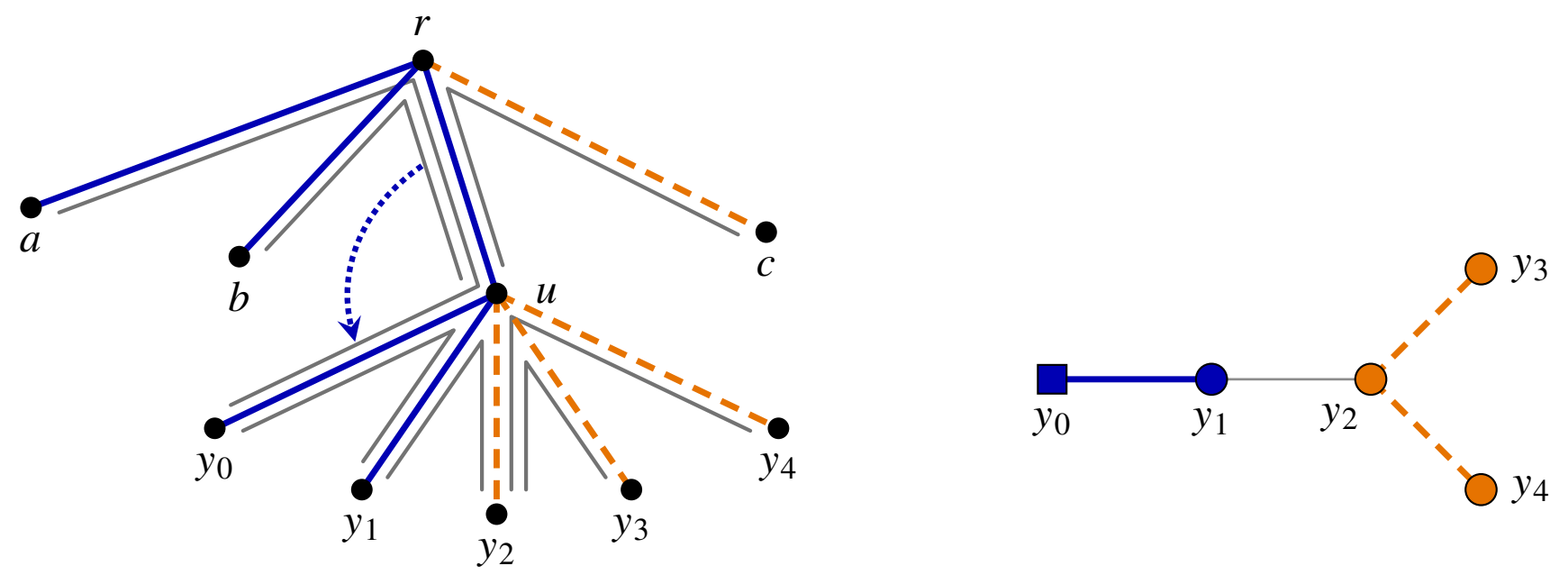
\begin{figure}[t]
\centering
\begin{tikzpicture}[xscale = 0.9, yscale = 0.9,
  vtx/.style={circle, fill=black, inner sep=0pt, minimum size=4.5pt},
  c1/.style={colOne, line width=1.3pt},
  c2/.style={colTwo, line width=1.3pt, dash pattern=on 4pt off 2.5pt},
  inherit/.style={-stealth, colOne, line width=0.6pt, very thick, densely dotted},
  commodity/.style={black!55, line width=0.5pt, line cap=round, line join=round, thick},
]
  \foreach \n/\p in {r/{0,0}, a/{-3.2,-1.2}, b/{-1.5,-1.6}, w/{0.6,-1.9}, c/{2.8,-1.4}, y0/{-1.7,-3}, y1/{-0.5,-3.5}, y2/{0.6,-3.7}, y3/{1.7,-3.5}, y4/{2.9,-3}}
    \coordinate (\n) at (\p);

  \draw[commodity] (-2.97,-1.24) -- (-0.08,-0.16) -- (0.45,-1.84) -- (-1.57,-2.81);
  \draw[commodity] (-1.28,-1.54) -- (-0.15,-0.33) -- (0.31,-1.78);
  \draw[commodity] (2.57,-1.42) -- (0.20,-0.23) -- (0.65,-1.67);
  \draw[commodity] (-1.47,-3.02) -- (0.25,-2.20) -- (-0.49,-3.27);
  \draw[commodity] (-0.29,-3.40) -- (0.48,-2.29) -- (0.48,-3.50);
  \draw[commodity] (0.84,-3.50) -- (0.84,-2.46) -- (1.49,-3.40);
  \draw[commodity] (0.72,-3.50) -- (0.72,-2.09) -- (2.67,-3.02);
  \draw[c1] (r) -- (a);  \draw[c1] (r) -- (b);  \draw[c1] (r) -- (w);  \draw[c2] (r) -- (c);
  \draw[c1] (w) -- (y0); \draw[c1] (w) -- (y1); \draw[c2] (w) -- (y2);
  \draw[c2] (w) -- (y3); \draw[c2] (w) -- (y4);
  \foreach \n in {r,a,b,w,c,y0,y1,y2,y3,y4} \node[vtx] at (\n) {};
  \node[above=2pt] at (r) {$r$};
  \node[below=2pt] at (a) {$a$};
  \node[below=2pt] at (b) {$b$};
  \node[right=5pt] at (w) {$u$};
  \node[below=2pt] at (c) {$c$};
  \node[below=2pt] at (y0) {$y_0$};
  \foreach \i in {1,...,4} \node[below=2pt] at (y\i) {$y_\i$};
  \draw[inherit] ($(r)!0.4!(w)+(-.25,-0.1)$) to[bend right=35] ($(w)!0.45!(y0)+(-0.12,0.1)$);
  \node at (0.3,-4.6) {(a) Coloring of the tree $T$};
\end{tikzpicture}%
\hspace{1.5cm}%
\begin{tikzpicture}[xscale=0.9, yscale = 0.9,
  c1/.style={colOne, line width=1.3pt},
  c2/.style={colTwo, line width=1.3pt, dash pattern=on 4pt off 2.5pt},
  hv/.style={circle, draw=black, inner sep=0pt, minimum size=6.5pt},
]
  \coordinate (h0) at (-1.3,-2.6);
  \coordinate (h1) at (0,-2.6);
  \coordinate (h2) at (1.3,-2.6);
  \coordinate (h3) at (2.2,-1.7);
  \coordinate (h4) at (2.2,-3.5);
  \draw[black!45, line width=0.5pt] (h1) -- (h2);
  \draw[c2] (h2) -- (h3); 
  \draw[c2] (h2) -- (h4);
  \draw[c1] (h0) -- (h1);
  \node[draw=black, fill=colOne, inner sep=0pt, minimum size=6pt] at (h0) {};
  \foreach \n/\f in {h1/colOne, h2/colTwo, h3/colTwo, h4/colTwo} \node[hv, fill=\f] at (\n) {};
  \node[below=3pt] at (h0) {$y_0$};
  \node[below=3pt] at (h1) {$y_1$};
  \node[below left=2pt and -1pt] at (h2) {$y_2$};
  \node[right=3pt] at (h3) {$y_3$};
  \node[right=3pt] at (h4) {$y_4$};
  \node at (0.45,-4.6) {(b) Link graph $H_u$};
\end{tikzpicture}
\vspace*{-0.2cm}
\caption{A tree rooted at $r$ is given in (a); every commodity is a link and is drawn as a thin
line next to its edges. We have $\conge = 3$ which gives $q=2$, and the two edge color classes are indicated by blue, solid edges and orange, dashed edges. After coloring the children of $r$, we see that there is a bent commodity including both $\{r,u\}$ and $\{u, y_0\}$, and thus the set of pinned vertices at $u$ is $Z_u = \{y_0\}$ and $\{u,y_0\}$ inherits the color of its parent edge (indicated by the dotted arrow). In (b), we show the link graph $H_u$ where the pinned child of $u$ is shown as a square. We apply \Cref{l.Lovasz} to $H_u - Z_u$ which yields the vertex coloring shown in (b) and in turn induces the coloring of the child edges of $u$ shown in (a).}
\label{fig:construction_colouring}
\end{figure}
\newpage
\begin{restatable}{observation}{propertiesColoring}
\label{obs:strcture_coloring}
Consider the coloring $c: E \rightarrow [q]$.
For a vertex $v$ and a color $j \in [q]$ denote by $N_v^j$ the set of children $z$ of $v$ for which the edge $\{v,z\}$ has color $j$. Then,
\begin{enumerate}
    \item[(i)] Every side of every bent commodity is monochromatic,
    \item[(ii)] The graph $H_v[N_v^j \setminus Z_v]$ has maximum degree at most 2.
    \item[(iii)] For every vertex $v$, we have $|Z_v| \leq \conge$ and all pinned edges connecting $v$ to one of its children have the same color as $e_v$.
\end{enumerate}
\end{restatable}

\subsection{Solving the Color Classes}
\label{app:solvingColorClasses}

Fix a color $j \in [q]$, and recall that the feasible solutions to $\I_j$ are exactly those solutions to the original instance $\I$ which are contained in $E_j \coloneqq c^{-1}(j)$. 
To obtain the $q$-approximation of \Cref{thm:congestion_approx_halldorsson_lau}, it suffices to show that we can solve each $\I_j$ optimally since every feasible solution to $\I_j$ is also feasible for $\I$. In this subsection, we present a dynamic program which makes use of the structure of the computed coloring $c$ (cf. \Cref{obs:strcture_coloring}) to solve each of the color classes to optimality.

We begin by introducing further notation. For a vertex $v$, let $S_v$ be the set of commodities containing its parent edge, and denote by $E_v^{\downarrow}$ the edge set consisting of $e_v$ and all edges below it. For the root $r$ let $S_r \coloneqq \emptyset$ and $E_r^{\downarrow} \coloneqq E$. We say that a link at $v$ is \emph{active} if both of its sides contain an edge of color $j$, and \emph{inactive} otherwise. \\

\noindent\textbf{Table entries.} We compute table entries $f_v(a)$ for every vertex $v \neq r$ and every \emph{budget vector} $a = (a_i)_{i \in S_v}$ with $a_i \in \{0, \ldots, b_i\}$ denoting the remaining budget on commodity $i$. Since $|S_v| \leq \conge$, there are at most $n \cdot (b_{\max} + 1)^{\conge}$ table entries. We define the values $f_v(a)$ as
\[f_v(a) \coloneqq \max \left\{w(F) \ \big| \ F \subseteq E_v^\downarrow \cap E_j, \ |F \cap P_i| \leq a_i \ \forall i \in S_v, \ |F \cap P_i| \leq b_i \ \forall i \not\in S_v\right\}.\]
That is, $f_v(a)$ describes the maximum objective value that can be obtained when restricted to edges of color $j$ that are below $e_v$ in the tree, and when adhering to the remaining budgets specified by the budget vector $a$ for all commodities in $S_v$. \\

\noindent\textbf{Recursive formula.} We give a recurrence to compute the values $f_v(a)$. This recurrence holds for every edge coloring $c: E \rightarrow [q]$, however, the properties of \cref{obs:strcture_coloring} are needed to ensure that it can be evaluated efficiently. Fix a current vertex $v$. The main idea of our recurrence is that for each active link $P_i$ we choose a \emph{split} $\sigma$ which designates how much budget we assign to each of the two sides of $P_i$. More formally, if the active link $P_i$ at $v$ includes the children $z$ and $z'$ of $v$, we choose nonnegative integers $\sigma_i(z)$ and $\sigma_i(z')$ satisfying $\sigma_i(z) + \sigma_i(z') = b_i$. Given a budget vector $a$ at $v$ and a split $\sigma$, every child $z$ of $v$ receives the budget vector $a_z^\sigma$ defined by
\[(a_z^\sigma)_i \coloneqq \begin{cases}
    a_i, & i \in S_v, \\
    \sigma_i(z), & P_i \text{ is an active link at }v, \\
    b_i, & \text{otherwise, i.e., }P_i\text{ is vertical or inactive}
\end{cases}\]
for all $i \in S_z$. Note that inactive links at $v$ receive full budget on both of their sides. This is harmless, since at most one of the two sides can include an edge of color $j$. We maximize over all splits and define
\[\Phi_v(a) \coloneqq \max_{\sigma} \sum_z f_z(a_z^\sigma).\]
With this, we can give a recursive formula for $f_v(a)$.

\begin{lemma}
\label{lem:DP_recursion}
    Let $c: E \rightarrow [q]$ be arbitrary. For every vertex $v \neq r$ and every budget vector $a$ at $v$, we have
    \[f_v(a) = \begin{cases}
        \max\{\Phi_v(a), w_{e_v} + \Phi_v(a - \bf{1})\}, & e_v \in E_j \text{ and }a_i \geq 1 \text{ for all }i \in S_v, \\
        \Phi_v(a), & \text{otherwise.}
    \end{cases}\]
    Moreover, the optimal objective value that can be obtained from instance $\I_j$ is $\Phi_r(\emptyset)$.
\end{lemma}

\begin{proof}
    Let $v$ be a vertex, $a$ be a budget vector at $v$, and let $F \subseteq E_j$ be a set of edges below $v$. For a child $z$ of $v$, let $F_z \coloneqq F \cap E_z^\downarrow$. We show that the conditions $|F \cap P_i| \leq a_i$ for all $i \in S_v$ and $|F \cap P_i| \leq b_i$ for all $i \not\in S_v$ from the definition of the table entries $f_v(a)$ hold if and only if there exists a split $\sigma$ such that, for every child $z$, the set $F_z$ satisfies the constraints of $f_z(a_z^\sigma)$. We refer to this equivalence by $(\star)$. To show this, we only need to consider commodities $P_i$ which include at least one edge below $v$. We distinguish four cases.
    \begin{itemize}
        \item If $P_i$ contains no edge from $v$ to one of its children, then it must lie entirely below some child $z$, and it is part of neither $S_v$ nor any sets $S_{z'}$. Thus, both conditions require $|F \cap P_i| = |F_{z} \cap P_i| \leq b_i$.
        \item If $P_i$ contains exactly one edge $\{v,z\}$ from $v$ to one of its children, we have $F \cap P_i = F_z \cap P_i$, and for both we need $|F \cap P_i| \leq a_i$.
        \item If $P_i$ is vertical or an inactive link at $v$, then $F \cap P_i$ can include edges below at most one child $z$ of $v$. For vertical commodities this is clear, and for inactive links, at most one of the two sides can include an edge of color $j$. Thus, both conditions require $|F \cap P_i| = |F_z \cap P_i| \leq b_i$.
        \item Finally, we consider the case where $P_i$ is an active link at $v$ with sides including the children $z$ and $z'$ of $v$. Clearly, $|F \cap P_i| = |F_z \cap P_i| + |F_{z'} \cap P_i|$. This is at most $b_i$ if and only if there exist $\sigma_i(z) + \sigma_i(z') = b_i$ with $|F_ \cap P_i| \leq \sigma_i(z)$ and $|F_{z'} \cap P_i| \leq \sigma_i(z')$.
    \end{itemize}
    This proves the claimed equivalence. For a fixed split $\sigma$, the constraints on the sets $F_z$ are independent of each other and we have $w(F) = \sum_z w(F_z)$. Hence, the maximum of $w(F)$ over all sets $F \subseteq E_j$ below $v$ satisfying the budget constraints given by $a$ is $\Phi_v(a)$.

    At the root $r$, we have $S_r = \emptyset$, and thus we only need to consider the empty budget vector. Hence, the optimal objective value of $\I_j$ is $\Phi_r(\emptyset)$. For $v \neq r$, let $a$ be a budget vector at $v$ and $F \subseteq E_v^\downarrow  \cap E_j$. If $e_v \not\in F$, then we can use the equivalence $(\star)$ proven above, which gives the first case of the two cases in the maximum expression in the lemma. Otherwise, we have $e_v \in F \cap E_j$. Since all commodities in $S_v$ contain $e_v$, the set $F$ satisfies the constraints of $f_v(a)$ if and only if $a_i \geq 1$ for all $i \in S_v$ and $F \setminus \{e_v\}$ satisfies the constraint of equivalence $(\star)$ for budget vector $a - \bf{1}$. This gives the second entry in the maximum. \qedhere
\end{proof}

Next, we show that we can use the structure of the coloring described in \Cref{obs:strcture_coloring} to efficiently compute the values of $\Phi_z(a)$.

\begin{lemma}
\label{lem:DP_efficiency}
    Let $v$ be a vertex, and suppose that all table entries $f_z(a)$ for all children $z$ of $v$ are known. Then, we can compute the values of $\Phi_v(a)$ for all budget vectors $a$ at $v$ in time $(b_{\max} + 1)^{\conge(|Z_v| + 4)} \cdot n^{\mathcal{O}(1)}$.
\end{lemma}

\begin{proof}
    There are at most $(b_{\max} + 1)^{\conge}$ budget vectors at $v$. We choose a fixed budget vector $a$. When choosing a split  $\sigma$ at $v$ we need to choose the value $\sigma_i(z) \in \{0, \ldots, b_i\}$ for every active link at $v$. Note that the value of $f_z(a_z^\sigma)$ for some child $z$ only depends on the choices made for active links through $z$. Enumerating all splits is too slow since the number of active links at $v$ is not bounded in terms of $\conge$. Instead, we process children one by one and keep track of all splits of links that join a processed child to an unprocessed one. Using \Cref{obs:strcture_coloring} we can ensure that (when choosing a suitable order of the children $z$) there are few such links.

    Let $P_i$ be an active link at $v$. As $P_i$ is active, both of its sides contain at least one edge of color $j$, and thus, by \Cref{obs:strcture_coloring}(i) also the two top edges of $P_i$ have color $j$. Hence, every active link at $v$ joins two children in $N_v^j$ which are adjacent in $H_v$. Therefore, for a child $z \not\in N_v^j$, the vector $a_z \coloneqq a_z^\sigma$ does not depend on $\sigma$. Thus, we have
    \begin{equation}
    \label{eq:relevant_parts_of_split}
    \Phi_v(a) = \sum_{z \not\in N_v^j} f_z(a_z) + \max_\sigma \sum_{z \in N_v^j} f_z(a_z^\sigma).
    \end{equation}
    Next we construct a good ordering of the children of color $j$, that is of $N_v^j \coloneqq \{z_1, \ldots, z_m\}$ for which there are few links joining a processed child to an unprocessed one. To this end, for a fixed ordering $(z_1, \ldots, z_m)$ and for $0 \leq t \leq m$, let $K_t$ be the set of active links at $v$ with one side containing a child in $\{z_1, \ldots, z_t\}$ and the other side containing a child in $\{z_{t+1}, \ldots, z_m\}$. By \Cref{obs:strcture_coloring}(ii), every connected component of $H_v[N_v^j \setminus Z_v]$ is a path or cycle. We construct the ordering of the children as follows:
    \begin{itemize}
        \item First, process the vertices $N_v^j \cap Z_v$ that correspond to pinned edges of color $j$.
        \item Then, process the components of $H_v[N_v^j \setminus Z_v]$ one by one, each one in order of its path or cycle.
    \end{itemize}
    We claim that this ordering fulfills
    \begin{equation}
        \label{eq:small_crossing}
        |K_t| \leq C(|N_v^j \cap Z_v| + 2) \qquad \text{for all }0 \leq t \leq m.
    \end{equation}
    Fix a child $z$. Clearly, all active links through $z$ contain the edge $\{v,z\}$ and thus there are at most $\conge$ such links. In particular, at most $C|N_v^j \cap Z_v|$ links in $K_t$ have a side through a child in $Z_v$, and every other link in $K_t$ joins two children $z_s, z_{s'} \not \in Z_v$ with $s \leq t < s'$. As these children are adjacent in $H_v[N_v^j \setminus Z_v]$ they must lie in a common component which is listed in path or cycle order in the ordering of the children $(z_1, \ldots, z_m)$. There can be at most two edges joining a vertex in $\{z_1, \ldots, z_t\}$ to a vertex in $\{z_{t+1}, \ldots, z_m\}$, those being $\{z_t, z_{t+1}\}$ and (if the component is a cycle) the edge between its last and first vertex. Each of these two edges accounts for at most $\conge$ links, and thus we have $|K_t| \leq C(|N_v^j \cap Z_v| + 2)$ which proves \eqref{eq:small_crossing}. If $j \neq c(e_v)$, \Cref{obs:strcture_coloring}(iii) shows $N_v^j \cap Z_v = \emptyset$ which implies the stronger bound $|K_t| \leq 2\conge$.

    We now process $N_v^j$ in the order $(z_1, \ldots, z_m)$ constructed above. For $0 \leq t \leq m$ we fix the values of the split $\sigma$ for all links in $K_t$ to be some vector $\tau$ and define
    \[\Psi_t(\tau) \coloneqq \max\left\{\sum_{s=1}^t f_{z_s}(a_{z_s}^\sigma) \ \bigg| \ \sigma \text{ is a split at } v \text{ with } \sigma|_{K_t} = \tau\right\}.\]
    That is, $\Psi_{t}(\tau)$ is the maximum weight that can be obtained from the first $t$ children of $v$ of color $j$ in the computed ordering if the budgets of all links in $K_t$ are split according to $\tau$. We have $K_0 = K_m = \emptyset$, and thus $\Psi_0(\emptyset) = 0$, and $\Psi_m(\emptyset)$ is the maximum value from \eqref{eq:relevant_parts_of_split}. For an active link through $z_t$ there are two options: Either it joins $z_t$ to an earlier child in which case it lies in $K_{t-1}$, or it joins $z_t$ to a later child, in which case it lies in $K_t$. Thus, $a_{z_t}^\sigma$ is uniquely defined through $\sigma|_{K_{t-1} \cup K_t}$ and hence,
    \[\Psi_t(\tau) = \max \left\{\Psi_{t-1}(\sigma|_{K_{t-1}}) + f_{z_t}(a_{z_t}^\sigma) \ \Big| \ \sigma \text{ is a split of } K_{t-1}\cup K_t \text { with } \sigma|_{K_t} = \tau\right\}.\]
    We iterate over all $t$, and for each fixed $t$ maximize over all possible splits $\sigma$ of $K_{t-1} \cup K_{t}$. This way, we can compute $\Psi_m(\emptyset)$ which maximizes \eqref{eq:relevant_parts_of_split}.

    We show that our approach yields the desired running time. The construction of the order $(z_1, \ldots, z_m)$ is possible in time $\mathcal{O}(n\conge)$. We have $|K_{t-1} \cup K_t| \leq |K_{t-1}| + \conge \leq \conge(|Z_v| + 3)$. Hence, for every $t$ we need to consider at most $(b_{\max} + 1)^{C(|Z_v|+3)}$ splits $\sigma$, each of which can be handled in time $\conge^{\mathcal{O}(1)}$. By summing over all $m \leq n$ children of $v$ and all at most $(b_{\max}+1)^{\conge}$ budget vectors at $v$ we get the desired runtime. \qedhere
\end{proof}

With this, we can solve each color class exactly. We conclude the proof of \Cref{thm:congestion_approx_halldorsson_lau}.

\begin{proof}[of \Cref{thm:congestion_approx_halldorsson_lau}]
The construction of the coloring only takes polynomial time. Now, consider each color class $j$ separately. In a preprocessing step, iterate through all commodities and determine whether they are active at their apex. This takes time $n^{\mathcal{O}(1)}$. Now, process vertices bottom-up using the dynamic program described in Lemmata \ref{lem:DP_recursion} and \ref{lem:DP_efficiency}. By these lemmata our DP works correctly and the iteration for some vertex $v$ takes time $(b_{\max}+1)^{\conge(\conge+4)}\cdot n^{\mathcal{O}(1)}$. By \Cref{obs:strcture_coloring}(iii), we have $|Z_v| \leq \conge$ for all $v \in V$, and thus, after iterating over all vertices $v$ obtain a running time of $(b_{\max}+1)^{\conge(|Z_v|+4)}\cdot n^{\mathcal{O}(1)}$ per color class.

After computing one candidate solution $F_j$ for each color class $j \in [q]$, we return the heaviest among these solutions. This solution is feasible for $\I$, and is in fact a $\lceil (\conge+1)/3\rceil$-approximation by \eqref{eq:coloring_approximation}. \qedhere
\end{proof}

\subsection{Solving the Color Classes is $\NP$-hard}
\label{app:colorClassesHardness}

Consider $\BPPT$ restricted to trees in which we can choose a root $r$ such that each vertex has at most 2 children with respect to root $r$. When computing the coloring $c: E \rightarrow [q]$ on such a graph, the link graph $H_v$ (and consequently also $H_v - Z_v$) has maximum degree at most $1$. Hence, the computed coloring of $E$ is monochromatic and we have $\I_1 = \I$.

We show that $\BPPT$ is $\NP$-hard when restricted to this class of graphs. This implies that there is no hope for a fully polynomial-time algorithm using the construction of the coloring presented in Appendix \ref{sec:constr_edge_coloring}.

\begin{observation}
$\BPPT$ is $\NP$-hard when restricted to trees for which one can choose a root $r$ such that every vertex has at most two children.
\end{observation}
\begin{proof}
    We reduce from Independent Set on a graph $G=(V,E)$ with $V = \{v_1, \ldots, v_n\}$. To construct an instance of $\BPPT$ on a tree $G'$, take a path $(w_1, \ldots, w_n)$ and assign weight $0$ to all edges $\{w_i, w_{i+1}\}$. Then, connect $v_i$ to $w_i$ using an edge $\{w_i, v_i\}$ of weight one. Finally, for each edge $\{v_i, v_j\} \in E$, add a unit-budget commodity whose path connects $v_i$ and $v_j$ in $G'$. Clearly, in the constructed instance rooted at $w_1$, every vertex has at most two children. Furthermore, the optimal objective value that can be obtained in the constructed instance of $\BPPT$ is exactly the independence number $\alpha(G)$. \qedhere
\end{proof}
\end{document}